\documentclass[12pt]{iopart}
\usepackage{iopams}
\usepackage{epsf}
\usepackage{cite}
\usepackage{graphicx}
\usepackage{longtable}

\begin{document}

\title{Ionization of Rydberg atoms by blackbody radiation} % Declares the document's title.

\author{I.I.~Beterov$^1$\footnote[3]{To whom
correspondence should be addressed (beterov@isp.nsc.ru)},
D.B.~Tretyakov$^1$, I.I.~Ryabtsev$^1$, V.M.~Entin$^1$,
A.~Ekers$^2$, N.N.~Bezuglov$^3$} % Declares the author's name.
\address{$^1$Institute of Semiconductor Physics, Pr.~Lavrentyeva 13, 630090
Novosibirsk, Russia}
\address{$^2$University of Latvia, Laser Centre, LV-1002 Riga, Latvia}
\address{$^3$St. Petersburg State University, Faculty of Physics,
198904 St.-Petersburg, Russia}

\begin{abstract}

We have studied ionization of alkali-metal Rydberg atoms by
blackbody radiation (BBR). The results of theoretical
calculations of ionization rates of Li, Na, K, Rb and Cs Rydberg
atoms are presented. The calculations have been performed for
\textit{nS}, \textit{nP} and \textit{nD} states
for principal quantum numbers
\textit{n}=8-65 at ambient temperatures of 77, 300
and 600~K. The calculations take into
account the contributions of BBR-induced redistribution of
population between Rydberg states prior to photoionization and
field ionization by extraction electric field pulses. The obtained
results show that these phenomena affect both the magnitude of
the measured ionization rates and their $n$-dependence.
A Cooper minimum for BBR-induced transitions between
bound Rydberg states of Li has been found. The calculated
ionization rates are compared with our earlier measurements of
BBR-induced ionization rates of Na \textit{nS} and \textit{nD}
Rydberg states with \textit{n}=8-20 at 300~K. A good agreement for
all states except \textit{nS} with $n>15$ is observed. Useful
analytical formulas for quick estimates of BBR ionization rates
of Rydberg atoms are presented. Application of BBR-induced
ionization signal to measurements of collisional ionization rates
is demonstrated.

\end{abstract}
 \maketitle

\tableofcontents

%================================================================================
%================================================================================
\section{INTRODUCTION}
%================================================================================
%================================================================================
Studies of blackbody radiation (BBR) started when Robert
Kirchhoff noticed that BBR is of great importance in physics.
The results of those studies facilitated the development of quantum
mechanics. Today, after more than hundred years since its discovery, the blackbody
radiation is important and interesting for researchers working in various
areas of physics. The studies of BBR have revealed a number of new effects, for example anisotropy in the cosmic background
radiation \cite{Anisotropy}.

The most straightforward way to observe the interaction of
BBR with matter relies on the use of atoms in highly excited Rydberg
states with the principal quantum number $n \gg 1$
\cite{GallagherCooke1979}. Rydberg atoms have many unique
properties including large geometric size $\sim $\textit{n}$^{2}$,
large radiative lifetimes $\sim $\textit{n}$^{3}$, large
polarizabilities $\sim $\textit{n}$^{7}$ and relatively low
frequencies of transitions between neighboring states $\sim
$\textit{n}$^{-3}$. Since the dipole moments of low-frequency
transitions between Rydberg states are very large, Rydberg atoms
are extremely sensitive to electromagnetic fields, including
the BBR. The studies of interaction of BBR
with Rydberg atoms were initiated by Gallagher and Cooke
in 1979 \cite{GallagherCooke1979}. The authors of that pioneering
work demonstrated that the influence of BBR
must be taken into account in lifetime measurements, spectroscopy,
and all other measurements where population of Rydberg states is
monitored.

In the 1980-s, the interaction of Rydberg atoms with blackbody radiation
was studied in various contexts. The attention was mainly focused on calculations and
measurements of lifetimes of Rydberg states
\cite{GallagherPotassium1979, CookeGallagher1980, BBRXeStebbins,
Spencer1981, SpencerKleppner1982,Theodosiou} and BBR-induced Stark shifts
\cite{GallagherCooke1979,FarleyWing}. However, only a few studies
considered ionization of Rydberg atoms by BBR.
Interaction of a Rydberg atom A(\textit{nL}) with the
principal quantum number \textit{n} and the orbital momentum
\textit{L} with BBR leads not only to transitions
to other bound states, but also to transitions to the continuum:

\begin{equation}
\label{eq1} \mathrm{A}\left( {nL} \right) + \hbar \omega _{BBR}
\to \mathrm{A}^{+} + e^{ -}.
\end{equation}

\noindent Here, $\hbar \omega _{BBR}$ is the energy of the absorbed
BBR photon, A$^{+}$ is an atomic ion and $e^-$ is a free electron
emitted due to the ionization.

The first study of ionization of Rydberg atoms was published by
Spencer et al. \cite{SpencerIon1982}. They calculated and measured
the dependence of the sodium 17\textit{D} BBR-induced
photoionization rate on the ambient temperature. This study was
followed by numerical calculations of BBR-induced ionization rates
of H and Na atoms for a wide range of principal quantum numbers by
Lehman \cite{Lehman}. In 1986, Burkhardt~et~al.
\cite{Burkhardt1986} studied collisional ionization of Na Rydberg
atoms. They concluded that BBR-induced ionization is the main
mechanism of the atomic ion production. They also noticed that
state-mixing collisions affect the dependences of the measured
ionization rates on the principal quantum number. After an almost
decade long pause the studies of interaction of Rydberg atoms with
BBR were resumed by Galvez et al., who investigated the multistep
transitions between Rydberg states caused by BBR \cite{Galvez1995}
and the BBR-induced resonances between Rydberg states of Na in
static electric fields \cite{Galvez1997}. A few years later, Hill
et al. published a paper discussing the influence of the applied
electric field on collisional and BBR-induced ionization rates of
potassium Rydberg atoms \cite{DunningHill2000}.

Ionization of Rydberg atoms by BBR returned to the focus of
researchers in 2000 in connection with the observation of
spontaneous evolution of ultracold Rydberg atoms with $n>30$ into
an ultracold plasma initiated by BBR \cite{Robinson2000}. The
creation of ultracold neutral plasma by laser photoionization of
laser-cooled xenon atoms was reported by Killian et al. in 1999
\cite{Killian1999}. Numerous studies of ultracold plasma followed
immediately \cite{Killian2001, LiNoele2004, Ultracold1,
Ultracold1b, Ultracold2, Ultracold3, Ultracold4b, Ultracold5,
Ultracold6, Ultracold7, Ultracold8}. The ultracold plasma is an
example of a strongly coupled plasma (the thermal energy of
particles is less than the Coulomb interaction energy), which is
substantially different from the ordinary high temperature
plasmas. Strongly coupled plasmas appear in astrophysical systems,
but are rather difficult to obtain in the laboratory.

The mechanism of spontaneous formation of ultracold plasma was
described in \cite{LiNoele2004}. After laser excitation, cold
Rydberg atoms are ionized by blackbody radiation and by collisions
with the hot Rydberg atoms. The produced electrons quickly leave
the volume of a magneto-optical trap, but the cold ions do not.
Then the macroscopic positive charge of the ions attracts and
traps the electrons, making them oscillate back and forth through
the cloud of cold Rydberg atoms. Collisions of electrons with the
remaining Rydberg atoms lead to their rapid avalanche ionization.
The electrons are thermalized to typical temperatures of tens of
K, making such type of plasma really ultracold. The energy balance
in the system is maintained by collisional depopulation of highly
excited Rydberg states.

The estimates of BBR ionization rates in \cite{Robinson2000} were
based on a simple analytical formula presented by Spencer et al.
\cite{SpencerIon1982}, which is a generalization of the results,
obtained for Na 17\textit{D} state. In the recent work
\cite{LiNoele2004} a simple approximation for photoionization
cross-sections was used to calculate the BBR-induced ionization rate.
The photoionization cross-section was expressed only through the
energy of the Rydberg state and the energy of the absorbed photon,
neglecting the specific properties of alkali-metal Rydberg states with
low orbital momenta. Therefore, an extended systematic study of
BBR-induced ionization of alkali-metal Rydberg atoms is required.

Another possible application of BBR-induced ionization is the
correct measurement of collisional ionization rates of Rydberg
atoms. In such experiments \cite{PaperI} a BBR ionization signal
can be used as a reference for the determination of collisional
ionization rate constants.

In the present work we discuss the mechanism of BBR-induced
ionization of alkali-metal Rydberg atoms with low orbital momenta under realistic
experimental conditions. The existing theoretical approaches are analyzed and
compared. The simplest (but often insufficient) way of considering the
BBR-induced ionization after the excitation of an atom \textit{A} to a given
\textit{nL} Rydberg state is to take into account only the direct
photoionization.

%===============================================================================================
%=======================*************FIGURE 1**********=========================================
%===============================================================================================

\begin{figure}
\begin{center}
\includegraphics[width=8cm]{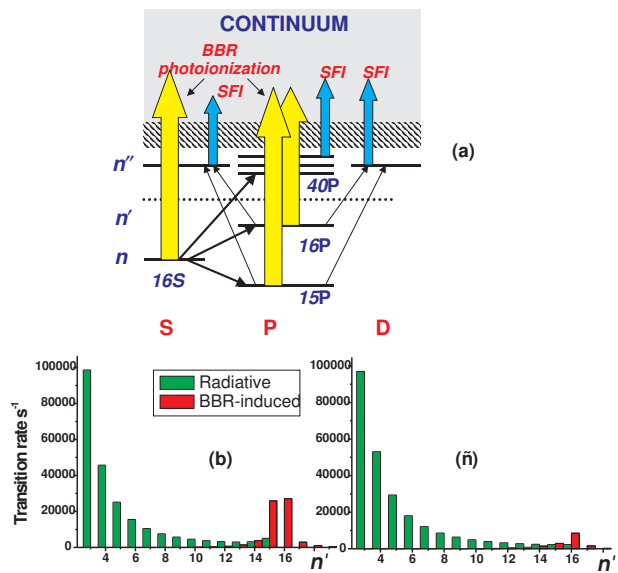}% Here is how to import EPS art
\caption{ \label{Fig1} (a) Schematic illustration of BBR-induced
and field ionization processes occurring after excitation of the
initial Na(16\textit{S}) state, including redistribution of
population over other $n'L'$ Rydberg states due to spontaneous and
BBR-induced transitions from the initial state. Highly excited
$n''S$, $n''P$ and $n''D$ Rydberg states are ionized by the
extracting electric pulses due to selective field
ionization~(SFI)\cite{SFI}. (b) Calculated spontaneous and
BBR-induced transition rates from the initial 16\textit{S} state
to other $n'P$ states. (c) Calculated spontaneous and BBR-induced
transition rates from the initial 16\textit{D} state to other
$n'P$ states.}

\end{center}
\end{figure}

%==============================================================================================

In the reality, however, ionization of Rydberg atoms exposed to
BBR is a complex process, in which the following main components
can be identified [see figure~\ref{Fig1}(a)]: (i) direct
photoionization of atoms from the initial Rydberg state via
absorption of BBR photons, (ii) selective field
ionization~(SFI)\cite{SFI} by extraction electric field pulses of
high Rydberg states, which are populated via absorption of BBR photons by atoms in the initial Rydberg
state, (iii) direct BBR-induced
photoionization of atoms from the neighboring Rydberg states,
which are populated due to absorption and emission of BBR photons
prior to the photoionization, and (iv) field ionization of other
high-lying states, which are populated via population
redistribution involving two or more steps of BBR photon
absorption and/or emission events. Our calculations show that all
these processes can contribute to the total ionization rate to a
comparable extent, and, therefore, none of them can be safely
disregarded. In Section~2 we will consider the above processes and
calculate the total BBR ionization rates, both analytically and
numerically.

We present the results of numerical calculation of BBR-induced
ionization rates for \textit{nS}, \textit{nP} and \textit{nD}
states of Li, Na, K, Rb and Cs atoms for a wide range of principal
quantum numbers \textit{n}=8-65 \cite{BBRIonization, JETP,
VestnikNSU}. We also present simple analytical formulas for quick
estimates of BBR-induced ionization rates. A Cooper minimum in the
discrete spectrum of Li will be discussed. Finally, the
theoretical results for Na \textit{nS} and \textit{nD} states are
compared with our experiment \cite{PaperI}.

All of the ionization mechanisms of Rydberg atoms exposed to BBR
are illustrated in figure~\ref{Fig1}. The total BRR-induced
ionization rate can be written as a sum of four separable
contributions:
\begin{equation}
\label{eq2}
W_{BBR}^{tot}=W_{BBR}+W_{SFI}+W_{BBR}^{mix}+W_{SFI}^{mix}.
\end{equation}

\noindent The first contribution, $W_{BBR} $, is the direct BBR
photoionization rate of the initially excited \textit{nL} state,
which will be discussed in subsection~2.1. The second term,
$W_{SFI} $, is the rate of SFI of high $n''L'$ Rydberg states,
which are populated from the initial Rydberg state \textit{nL} via
absorption of BBR photons. This field ionization is discussed in
subsection~2.3, while redistribution of population between Rydberg
states is described in subsection 2.2. The third term,
$W_{BBR}^{mix} $, is the total rate of BBR-induced photoionization
of neighboring $n'L'$ Rydberg states, which are populated via
spontaneous and BBR-induced transitions from the initial state.
The last term, $W_{SFI}^{mix} $, is the rate of SFI of high-lying
Rydberg $n''L'$ states that are populated in a two-step process
via absorption of BBR photons by atoms in $n'L'$ states (note that
here, in contrast to $W_{SFI} $, we consider lower $n'L'$, states
which cannot be field ionized). The two latter ionization rates,
which are related to population redistribution between Rydberg
states, are considered in subsection~2.4. The atomic units will be
used below, unless specified otherwise.

Experimental measurements of BBR-induced ionization rates are
discussed in Section~3. The temperature dependence of BBR-induced
ionization, measured by Spencer et al. \cite{SpencerIon1982}, is
discussed in subsection~3.1. The measured by us dependence of
the BBR-induced ionization rates of Na Rydberg states on the principal
quantum number \textit{n} is presented in subsection~3.2.
Application of the BBR-induced ionization to the measurements of collisional
ionization rates is discussed in subsection~3.3. The role of BBR
in the formation of ultracold plasma is reviewed in subsection~3.4.
Finally, the results of the present study are summarized in the
Conclusion.
%==========================================================================
%==========================================================================
%==========================================================================

\section{BBR-INDUCED IONIZATION: THEORETICAL APPROACH}

%==========================================================================

\subsection{Bound-bound transitions induced by blackbody
radiation}
%==========================================================================

We start the discussion of ionization of Rydberg atoms by
BBR with the consideration of BBR-induced
bound-bound transitions between Rydberg states, which have been studied
most extensively. Blackbody radiation causes both transitions
between Rydberg states and ac Stark shifts of energy levels
\cite{RydbergAtomsBook, RydbergAtomsNew, GallagherReview,
Filipovicz}. Large dipole moments of Rydberg states make them
sensitive to BBR. In addition, the spectral brightness
of BBR at \textit{T}=300~K (maximum at $2 \times
10^{13}$ Hz) is relatively high for the low frequencies of
transitions between Rydberg states.

Absorption of BBR by Rydberg atoms rapidly
redistributes the initial population to neighboring states and thus
reduces the selectivity of laser excitation. In
contrast to the spontaneous decay of Rydberg atoms, which populates
mostly ground and lower excited levels, the BBR-induced transitions
populate predominantly the neighboring Rydberg states [see
figure~\ref{Fig1}(b)]. Redistribution of the population of Rydberg
states by BBR can be suppressed by surrounding the laser
excitation volume with cooled shields. However, in order to reduce
the rate of BBR-induced transitions by an order of magnitude, a
liquid-helium cooling must be used.

Probabilities of BBR-induced transitions are proportional to the
number of photons per mode of the blackbody radiation
field\cite{RydbergAtomsBook}:

\begin{equation}
\label{eq3} \bar {n}_{\omega}  = \frac{{1}}{{\mathrm{exp}\left(
{\omega /kT} \right) - 1}},
\end{equation}
\noindent where \textit{kT} is the thermal energy in atomic units.
For atoms in the ground and low excited states with large
frequencies of transitions one has  $\bar {n}_{\omega} \gg 1$ at
\textit{T}=300~K, and the rates of BBR-induced transitions are
small. Hence, for atoms in those states interaction with BBR can
be disregarded. The situation is different for Rydberg states: at
transition frequencies on the order of 10~cm$^{-1}$ we have
$\bar{n}_{\omega}\sim 10$, and the rate of BBR-induced transitions
can be ten times larger than the rate of the spontaneous decay to
the neighboring Rydberg states.

Probability of a spontaneous transition between atomic $nL$ and
${n}'{L}'$ levels is given by the Einstein coefficient $A\left(
{nL \to n'L'} \right)$:

\begin{equation}
\label{eq4} A\left( {nL \to n'L'} \right) = \frac{{4
\omega_{nn'}^{3} }}{{3c^{3}}}\frac{{L_{max}} }{{2L +
1}}R^{2}\left( {nL \to n'L'} \right).
\end{equation}

\noindent Here, \textit{ L}$_{max}$ is the largest of $L$ and $L'$,
and $R\left( {nL \to n'L'} \right)$ is the radial matrix element of the
electric dipole moment. The total rate of the spontaneous decay is a
sum of rates of transitions to all states with $n' < n$:

\begin{equation}
\label{eq5} \Gamma _{nr} = \sum\limits_{L' = L \pm 1}
{\sum\limits_{n'<E_n}^{n} {A\left( {nL \to n'L'} \right)}}.
\end{equation}

\noindent The rate of BBR-induced transitions $W\left( {nL \to
n'L'} \right)$ between the states \textit{nL} and ${n}'{L}'$ is
given by $A\left( {nL \to n'L'} \right)$ and the number
of photons per mode of BBR at the transition
frequency $\omega _{nn'}=1/(2n^2)-1/(2n'^2)$:

\begin{equation}
\label{eq6} W\left( {nL \to n'L'} \right) = A\left( {nL \to
{n}'{L}'} \right)\frac{{1} }{{\mathrm{exp}\left(\omega_{nn'} /kT
\right) - 1}}.
\end{equation}

\noindent In contrast to spontaneous decay, blackbody radiation
populates states with the energy both higher and lower than that of the
initially excited state. The total rate of BBR-induced transitions
is a sum of rates of BBR-induced transitions to all $n'L'$ states:

\begin{equation}
\label{eq7} \Gamma _{BBR} = \sum\limits_{L' = L \pm 1}
{\sum\limits_{{n}'} {\Gamma _{BBR} \left( {n,L \to n',L'}
\right)}}.
\end{equation}

\noindent Blackbody radiation populates mostly the neighboring
states with ${n}' = n \pm 1$, which give the main contribution to
the total rate of BBR-induced transitions. This contribution to
the total decay rate of the initially excited Rydberg state can be
significant. The effective lifetime is inverse of the sum of total
decay rates due to spontaneous and BBR-induced
transitions:

\begin{equation}
\label{eq8} \tau _{eff}^{ - 1} = \Gamma _{nr} + \Gamma _{BBR}.
\end{equation}

\noindent The first experimental observation of depletion of
Rydberg atoms due to BBR by Gallagher and Cooke
\cite{GallagherCooke1979} was based on measurements of the
effective lifetimes of the 17\textit{P} and 18\textit{P}
states of sodium.

\begin{figure}
\begin{center}
\includegraphics[width=4cm]{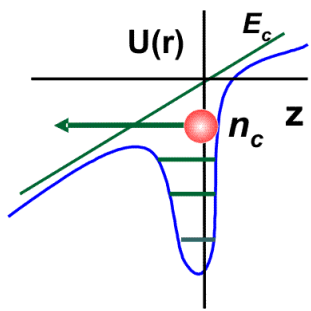}% Here is how to import EPS art
\caption{\label{Fig2}
Illustration of the selective field ionization (SFI) of a Rydberg
atom with a principal quantum number \textit{n}$_{c}$ by electric
field with the amplitude $E_c$.}
\end{center}

\end{figure}

A method to determine the effective lifetimes of Rydberg states
was discussed by Gallagher in Ref. \cite{RydbergAtomsBook}.
According to his calculations, the radiative lifetime of the
sodium 18\textit{S} state is 6.37~$\mu$s, but it reduces to
4.87~$\mu$s due to the interaction with BBR at \textit{T}=300~K. This
is consistent with
time-resolved measurements of the fluorescence signal on the $18S
\to 3P$ transition. The effective lifetime of the 18\textit{S}
state was also determined using the method of SFI \cite{SFI},
which is the only reliable way to measure the population of a
Rydberg state. Any Rydberg atom ionizes with a probability close
to unity if the electric-field strength has reached a critical
value $E_c$ (see figure~\ref{Fig2}). The latter strongly depends
on the effective quantum number $n_{eff} = n - \mu_L $, where
$\mu_L$ is the quantum defect:

\begin{equation}
\label{eq9} E_{c} \approx 3.2 \cdot 10^{8}n_{eff}^{ - 4} \quad
\left( \mathrm{V/cm} \right).
\end{equation}

\noindent Unfortunately, SFI is difficult to use for
Rydberg states with low \textit{n}, since it requires very strong
electric fields ($\sim $30~kV/cm for $n\sim $10).

%===============================================================================================
%=======================*************FIGURE 2**********=========================================
%===============================================================================================

%==============================================================================================

Haroche et al. \cite{Haroche1979} observed that the decay of the
25\textit{S} state of Na is accompanied by population of
the neighboring states including the 25\textit{P} state, which could be
populated only due to BBR. Later, in a well-known paper
\cite{Theodosiou} Theodosiou published the
calculated effective lifetimes of \textit{nS}, \textit{nP},
\textit{nD}, and \textit{nF} Rydberg states of alkali-metal atoms with $n<20$
at the ambient temperatures of 0, 350, 410, and 600~K. These values were obtained using the
accurate and reliable method of model potential.

Redistribution of population of Xe Rydberg states by BBR was
studied by Beiting et al. \cite{Beiting} using the SFI method.
The results of numeric calculations of depopulation rates of \textit{nS},
\textit{nP} and \textit{nD} ($n<30$) Rydberg states induced by
BBR were published in \cite{FarleyWing}. Radial
matrix elements of electric dipole transitions for states with
$n<15$ were calculated using the Bates-Damgaard method. For higher
Rydberg states the calculations were done in the Coulomb
approximation using the Numerov method \cite{Zimmerman}. The
populations of neighboring Rydberg states were calculated using a
one-step model, which took into account only the direct transitions from the
initially excited state. Suppose a
chosen \textit{nS} state is initially excited. The time-dependent
number of Rydberg atoms in this state is determined by its effective
lifetime $\tau _{eff}^{nS}$:

\begin{equation}
\label{eq10} N_{nS} \left( {t} \right) = N_{nS} \left( {0}
\right)\mathrm{exp}\left( { - {{t} \mathord{\left/ {\vphantom {{t}
{\tau _{eff}^{nS}} }} \right. \kern-\nulldelimiterspace} {\tau
_{eff}^{nS}} }} \right),
\end{equation}

\noindent where \textit{ N}$_{nS}$(0) is the number of Rydberg
atoms in the \textit{n}S state at time \textit{t}=0. The number of
atoms in the neighboring $n'P$ states is determined by two
competing processes: (\textit{i}) transitions [spontaneous at the rate
$A\left( {nS \to n'P} \right)$ and BBR-induced at the rate
$W\left( {nS \to n'P} \right) \sim $10$^{4}$ s$^{-1}$] from $nS$ states
with $n \sim 15$, and (\textit{ii}) spontaneous decay of $n'P$ states with
$n'\sim 15$ at the rate $\Gamma _{eff}^{n'P} (\sim
10^5\,\mathrm{s}^{-1})$:

\begin{equation}
\label{eq11} \frac{{dN_{n'P} \left( {t} \right)}}{{dt}} = \left[
{W\left( {nS \to n'P} \right) + A\left( {nS \to n'P} \right)}
\right]N_{nS} \left( {t} \right) - \Gamma _{eff}^{n'P} N_{n'P}
\left( {t} \right).
\end{equation}

\noindent The $n'P$ states are not populated initially. A solution
of equations~(\ref{eq10}) and (\ref{eq11}) with the initial
condition $N_{n'P} \left( {0} \right) = 0$ is:

\begin{eqnarray}
\label{eq12} N_{n'P} \left( {t} \right) = \frac{{\left[ {W\left(
{nS \to n'P} \right) + A\left( {nS \to n'P} \right)} \right]N_{nS}
\left( {0} \right)}}{{\Gamma _{eff}^{n'P} - \Gamma _{eff}^{nS}}
}\times &&\\\nonumber \qquad \qquad \qquad \qquad  \times\left(
{\mathrm{exp}\left( { - \Gamma _{eff}^{nS} t} \right) -
\mathrm{exp}\left( { - \Gamma _{eff}^{n'P} t} \right)} \right).&&
\end{eqnarray}

\noindent The range of applicability of the one-step model was
discussed by Galvez et al. \cite{Galvez1995}. They have developed
a multistep model and studied the redistribution of population
theoretically and experimentally. Rydberg atoms in a sodium atomic
beam were excited by two pulsed dye lasers. The number density of
ground-state atoms in the interaction region was varied from $8
\times 10^{9}\,\mathrm{cm}^{ - 3}$ to $2 \times
10^{10}\,\mathrm{cm}^{-3}$. At the time \textit{t}$_{d}$ after the
excitation, the populations of Rydberg states were detected using
the time-resolved SFI method \cite{SFI} by an electric filed pulse
with the amplitude of 1~kV and the duration of 4~$\mu$s, which was
sufficient to ionize all states with $n \ge 24$. When the delay
time $t_d$ was increased from 0 to 40~$\mu$s, additional peaks in
the field ionization spectrum appeared, indicating that
neighboring Rydberg states were populated by BBR.

The population $N_{i} $ of the \textit{i}th Rydberg state is a
solution of a system of differential equations \cite{Galvez1995}:

\begin{equation}
\label{eq13} \frac{{dN_{i}} }{{dt}} = - N_{i} \sum\limits_{j}
{\Gamma _{j}^{i} + \sum\limits_{k} {N_{k} \Gamma _{i}^{k}}},
\end{equation}

\noindent which take into account the multistep transitions. Here,
$\Gamma_j^i$ is the total rate of transitions between states \textit{i}
and \textit{j}. The first term in the rhs of
equation~(\ref{eq13}) describes the decay of state \textit{i}, while second term gives the population of state \textit{i}
due to the decay from higher states populated by BBR. The authors of
Ref. \cite{Galvez1995} solved a system of 32
equations~(\ref{eq13}) for (\textit{n}+1)\textit{S}, \textit{nP},
\textit{nD} and \textit{nF} states with \textit{n}=25-32. One-,
two- and three-step transitions were taken into account, while for
higher steps it was necessary to take into account the states with larger
orbital momenta. However, it was shown that the population of
\textit{nF} states was small, and the contribution from states with
larger \textit{L} can be neglected. Evolution of
Rydberg state populations was calculated numerically. Radial matrix elements
were calculated using the Van~Regermorter method
\cite{Regermorter}, which is fast and agrees with the more
complicated Numerov method at an accuracy level of 1\%. The results
showed that if the delay time \textit{t}$_{d}$ is comparable with the
lifetime of the initially excited state, the multi-step
transitions play an important role in the population redistribution,
which cannot be described using the one-step model. BBR-induced
resonances between Stark states of sodium Rydberg atoms have also
been studied \cite{Galvez1997}.

The above consideration of BBR-induced transitions between Rydberg
states was based on the electric dipole approximation and
perturbation theory. The range of applicability of such approach
was analyzed by Farley and Wing \cite{FarleyWing}. At the room
temperature \textit{T}=300~K, the energy of BBR photons is
comparable with the energy of Coulomb interaction of Rydberg
electron with the atomic core at $n\sim 120$, which is the limit
of applicability for the perturbation theory. The dipole approximation
breaks down when the wavelength of BBR is
comparable with the orbit size of the Rydberg electron ($n\sim 200$
at the room temperature).
%=================================================================================
%=================================================================================
%=================================================================================
\subsection{Direct BBR photoionization}
%=================================================================================
%=================================================================================
The direct BBR photoionization rate $W_{BBR} $ of a given \textit{nL}
state is calculated from the general formula
\cite{SpencerIon1982}:

\begin{equation}
\label{eq14} W_{BBR} = c\int\limits_{\omega _{nL}} ^{\infty}
{\sigma _{\omega}  \;\rho _{\omega}  d\omega},
\end{equation}

\noindent where \textit{c} is the speed of light, $\omega _{nL} =
1/\left( {2n_{eff}^{2}}  \right)$ is the photoionization threshold
frequency for \textit{nL} Rydberg state with the effective
principal quantum number \textit{n}$_{eff}$, and $\sigma _{\omega}
$ is the photoionization cross-section at frequency $\omega $. The
volume density $\rho _{\omega}  $ of BBR photons at temperature
\textit{T} is given by the Plank distribution:

\begin{equation}
\label{eq15} \rho_{\omega}  = \frac{{\omega^2}}{ {{\pi^2 c^3}
\left[ \mathrm{exp}\left(\omega /kT \right) -1 \right]}}.
\end{equation}

\noindent For isotropic and non-polarized thermal radiation field
the value of $\sigma _{\omega}  $ is determined by the radial
matrix elements $R\left( {nL \to E,L \pm 1} \right)$ of dipole
transitions from discrete \textit{nL} Rydberg states to the
continuum states with \textit{L$ \pm $}1 and photoelectron energy
\textit{E}:

\begin{equation}
\label{eq16} \sigma _{\omega}  = \frac{{4\pi^2\omega}}{{3c\left(
{2L + 1} \right)}}\sum\limits_{L' = L \pm 1} {L_{max} R^{2}\left(
{nL \to E,L \pm 1} \right)},
\end{equation}

\noindent where $L_{max}$ is the largest of $L$ and $L'$.

The main problem in the calculation of \textit{W}$_{BBR}$ for an arbitrary
Rydberg state is associated with finding $R\left( {nL \to E,L \pm 1} \right)$ and its
frequency dependence. In order to achieve a high accuracy of the matrix
elements, numerical calculations should be used.

Spencer et al. \cite{SpencerIon1982} studied the temperature dependence of the
rate of direct BBR-induced photoionization \textit{W}$_{BBR}$ of
the 17\textit{D} state of Na. The values of \textit{W}$_{BBR}$ were calculated
numerically and a simple formula was obtained:

\begin{equation}
\label{eq17} W_{BBR}\sim E_{n}^{2} \left[ {\mathrm{exp}\left(
{\frac{{E_{n}} }{{kT}}} \right) - 1} \right]^{ - 1},
\end{equation}

\noindent where \textit{E}$_{n}$ is the energy of the Rydberg
electron. This approximate formula was used for estimates of
\textit{W}$_{BBR}$ in many recent works on ultracold plasma
\cite{Robinson2000,LiNoele2004}. Accurate numerical calculations
of \textit{W}$_{BBR}$ using the method of model potential were
done by Lehman \cite{Lehman} for principal quantum numbers
\textit{n}=10-40 and temperatures \textit{T}=77-625~K, but only
for sodium and hydrogen atoms. Recently, the method of model
potential was used by Glukhov and Ovsiannikov \cite{Glukhov} to
calculate $W_{BBR}$ of helium \textit{nS}, \textit{nP} and
\textit{nD} Rydberg states. A simple analytical formula which
approximates the numerical results was obtained:

\begin{equation}
\label{eq18} W_{BBR} = \left( {a_{1} x^{2} + a_{2} x^{3} + a_{3}
x^{4}} \right)\frac{{1}}{{\mathrm{exp}\left( {x} \right) - 1}},
\quad x = \frac{{E_{n}} }{{kT}}.
\end{equation}

\noindent The coefficients $a_{1} ,a_{2} ,a_{3} $ depend on
the ambient temperature \textit{T}:

\begin{equation}
\label{eq19} a_{i} = \sum\limits_{k = 0}^{3} {b_{ik} \left(
{\frac{{T}}{{100}}} \right)^{k}}.
\end{equation}

\noindent The coefficients \textit{b}$_{ik}$, which depend only on
\textit{L}, were calculated independently for singlet and triplet
\textit{S}, \textit{P} and \textit{D} states of helium.

In present study we used the semi-classical formulas for dipole
matrix elements derived by Dyachkov and Pankratov \cite{Dyachkov1,
DyachkovPankratov}. In comparison with other semi-classical
methods \cite{GDK, DavydkinZon}, these formulas are advantageous
as they give orthogonal and normalized continuum wave functions,
which allow for the calculation of photoionization cross-sections
with high accuracy. We have verified that photoionization
cross-sections of the lower sodium \textit{S} states calculated
using the approach of \cite{DyachkovPankratov} are in good
agreement with the sophisticated quantum-mechanical calculations
by Aymar \cite{Aymar}.

A more accurate analytical expression for \textit{W}$_{BBR}$ than equation~(\ref{eq17}) is useful in order to illustrate the dependence of ionization rate on
\textit{n}, \textit{L}, and \textit{T}. We have obtained such
expression using the analytical formulas for bound-bound and
bound-free matrix elements derived by Goreslavsky, Delone, and
Krainov (GDK) \cite{GDK} in the quasiclassical approximation. For
the direct BBR photoionization of a \textit{nL} Rydberg state the
cross-section is given by:

\begin{equation}
\label{eq20} \sigma _{\omega}  \left( {nL \to E,L \pm 1} \right) =
\frac{{4L^{4}}}{{9cn^{3}\omega} }\left[ {K_{2/3} ^{2}\left(
{\frac{{\omega L^{3}}}{{3}}} \right) + K_{1/3} ^{2}\left(
{\frac{{\omega L^{3}}}{{3}}} \right)} \right],
\end{equation}

\noindent where $K_{\nu}  \left( {x} \right)$ is the modified
Bessel function of the second kind. This formula was initially
derived to describe the photoionization of hydrogen atoms.

The main contribution to \textit{W}$_{BBR}$ in
equation~(\ref{eq14}) comes from ionization by BBR of frequencies close to the
ionization threshold frequency $\omega _{nL} $, because the Plank
distribution rapidly decreases with increasing $\omega $. For
Rydberg states with \textit{n$ >  > $}1 and low \textit{L} one has
$\left( {{{\omega L^{3}} \mathord{\left/ {\vphantom {{\omega
L^{3}} {3}}} \right. \kern-\nulldelimiterspace} {3}}} \right) < <
1$, and equation~(\ref{eq20}) can be simplified to the
form:

\begin{equation}
\label{eq21}
\sigma _{\omega}  \left( {nL \to E,L \pm 1} \right) \approx
\frac{{1}}{{9cn^{3}}}\left[ {\frac{{6^{4/3}\Gamma ^{2}\left( {2/3}
\right)}}{{\omega ^{7/3}}} + \frac{{6^{2/3}\Gamma ^{2}\left( {1/3}
\right)}}{{\omega ^{5/3}}}L^{2}} \right].
\end{equation}

\noindent The combination of equations~(\ref{eq14}), (\ref{eq18})
and (\ref{eq20}) yields:

\begin{equation}
\label{eq22} W_{BBR} \approx \frac{{1}}{{\pi
^{2}c^{3}n^{3}}}\int\limits_{\omega _{nL} }^{\infty}  {\left[
{2.22\,\omega ^{ - 1/3} + 2.63\,\omega ^{1/3}L^{2}} \right]}
\frac{{d\omega} }{{\mathrm{exp}\left(\omega /kT \right) - 1}}.
\end{equation}

\noindent The expression in the square brackets is a slowly
varying function of $\omega $. Taking into account that the main
contribution to \textit{W}$_{BBR}$ is due to frequencies close to
the ionization threshold frequency, one can replace in the square
brackets $\omega $ by 1/(2\textit{n}$^{2}$). After such
replacement the integral in equation~(\ref{eq22}) can be
calculated analytically, and the final result is:

\begin{equation}
\label{eq23} W_{BBR} \approx \frac{{kT}}{{\pi ^{2}c^{3}}}\left[
{\frac{{2.80}}{{n^{7/3}}} + \frac{{2.09L^{2}}}{{n^{11/3}}}}
\right]\;\mathrm{ln}\left( {\frac{{1}}{{1 - \mathrm{exp}\left( { -
\frac{{\omega _{nL}} }{{kT}}} \right)}}} \right).
\end{equation}

\noindent Equation~(\ref{eq23}) gives the approximate direct
BBR-induced photoionization rate in atomic units for \textit{T}
measured in Kelvins.

In Ref.~\cite{GDK} it was proposed that equation~(\ref{eq20}) can
be extended to alkali-metal atoms simply by replacing \textit{n}
with $n_{eff} = \left( {n - \mu _{L}} \right)$. In the reality,
however, its accuracy is acceptable only for truly hydrogen-like
states with small quantum defects. A disadvantage of the GDK model
is that it disregards the non-hydrogenic phase factors in the
overlap integrals of dipole matrix elements. Nevertheless, we
suggest that for alkali-metal atoms equation~(\ref{eq23}) can be
rewritten as follows (for convenience \textit{W}$_{BBR}$ is
expressed in s$^{-1}$ for temperature \textit{T} taken in
Kelvins):

\begin{equation}
\label{eq24} W_{BBR} = C_{L} T\left[
{\frac{{14423}}{{n_{eff}^{7/3}} } +
\frac{{10770L^{2}}}{{n_{eff}^{11/3}} }} \right]\;\mathrm{ln}\left(
{\frac{{1}}{{1 - \mathrm{exp}\left( { -
\frac{{157890}}{{Tn_{eff}^{2}} }} \right)}}} \right)\quad
\,\,\left[ {\mathrm{s}^{ - 1}} \right].
\end{equation}

\noindent Here, \textit{C}$_{L}$ is an \textit{L}-dependent scaling
coefficient, which will be discussed below. A comparison of the
numerically calculated \textit{W}$_{BBR}$ with
equation~(\ref{eq24}) at \textit{C}$_{L}$=1 has shown a noticeable
disagreement of absolute values of \textit{W}$_{BBR}$, especially
for \textit{nS} states, which have large quantum defects (for
example, in sodium atoms the quantum defects are $\mu _{S}$=1.348,
$\mu_{P}$=0.855 and $\mu_{D}$=0.015). Formally, the disagreement
for the non-hydrogenic \textit{nS} states stems from peculiarities
of the asymptotic behavior of Bessel functions in
equation~(\ref{eq20}) for states with $L\ll1$. For example, the
analytical expression of GDK model yields close photoionization
cross-section values for the sodium \textit{nS}, \textit{nP} and
\textit{nD} states, while the accurate numerical calculations
yield significantly smaller cross-sections for the sodium
\textit{nS} states. At the same time, the shapes of the analytical
curves are quite similar to the numerical ones. Therefore, it is reasonable to introduce a scaling coefficient \textit{C}$_{L}$ in
equation~(\ref{eq24}) in order to make it valid for Rydberg states
of alkali-metal atoms with large quantum defects.

In fact, the scaling coefficient \textit{C}$_{L}$ accounts for
phase shifts of radial wave functions of alkali-metal Rydberg
states due to quantum defect. Delone, Goreslavsky and Krainov
\cite{GDK1989} suggested an approximate formula to calculate the
radial matrix elements of transitions between continuum states of
non-hydrogen atoms:

\begin{equation}
\label{eq25} R_{EL}^{E{l}'} \approx \frac{{0.4744}}{{\omega
^{5/3}}}\mathrm{cos}\left( {\Delta_{L} \pm \frac{{\pi} }{{6}}}
\right).
\end{equation}

\noindent Here, $\Delta_{L} = \left| {\pi \left( {\mu_{{L}'} -
\mu_{L}} \right)} \right|$ is the difference of quantum defects of
$L$ and $L'$ states, the $(+)$ sign corresponds to transitions
with ${L}' > L$ and the $(-)$ sign corresponds to transitions with
${L}' < L$. In order to take into account the phase shifts of
non-hydrogen wave functions, we have empirically introduced the
corrected non-hydrogen radial matrix elements:

\begin{equation}
\label{eq26} \tilde {R}_{nL}^{EL + 1} \sim R_{nL}^{EL + 1}
\mathrm{cos}\left( {\Delta_{L}^{ +}  + \frac{{\pi} }{{6}}}
\right), \quad \tilde {R}_{nL}^{EL - 1} \sim R_{nL}^{EL - 1}
\mathrm{cos}\left( {\Delta_{L}^{ -}  - \frac{{\pi} }{{6}}}
\right).
\end{equation}

\noindent Here, $\Delta_{L}^{ +}  = \pi \left( {\mu_{L} - \mu_{L +
1}} \right)$, $\Delta _{L}^{ -}  = \pi \left( {\mu_{L - 1} -
\mu_{L}} \right)$, and $R_{nL}^{EL \pm 1} $ are radial matrix
elements of bound-free transitions calculated in the hydrogen GDK
model with \textit{n} replaced by \textit{n}$_{eff}$. The
differences of quantum defects $\mu_{L} - \mu_{L'} $
\cite{QuantDefect,Hartree,LiDef,NaDef,KDef,RbDef,CsDef,Alkali1986}
for transitions from Rydberg states with \textit{n}$\sim $20 to
continuum are summarized in Table 1 for all alkali-metal atoms.

\newcommand{\PreserveBackslash}[1]{\let\temp=\\#1\let\\=\temp}
\let\PBS=\PreserveBackslash
\begin{longtable}
{|p{75pt}|p{90pt}|p{90pt}|p{90pt}|} a & a & a & a  \kill \hline
\multicolumn{4}{|p{335pt}|}{\textbf{Table 1}. Difference of
quantum defects of alkali-metal Rydberg states.}  \\ \hline &
 $\mu_{S} - \mu_{P} $ \textit{}&
 $\mu_{P} - \mu_{D} $ \textit{}&
 $\mu_{D} - \mu_{F} $ \textit{} \\
\hline Li& 0.352417& 0.0451664& 0.00162407 \\ \hline Na& 0.493519&
0.840023& 0.0148029 \\ \hline K& 0.466733& 1.43762& 0.264237 \\
\hline Rb& 0.490134& 1.29456& 1.34636 \\ \hline Cs& 0.458701&
1.12661& 2.43295 \\ \hline
\end{longtable}

In the calculations of photoionization rates for low-\textit{L}
states, the terms of equation~(\ref{eq20}) proportional to
\textit{L} and \textit{L}$^{2}$ can be neglected. Taking
into account equation~(\ref{eq25}), equation~(\ref{eq24}) can be
rescaled in order to achieve a better agreement with the numerical
results:

\begin{eqnarray}
\label{eq27} W_{BBR} = A_{L} \frac{{11500T}}{{n_{eff}^{7/3}}
}\left[ {\mathrm{cos}\left( {\Delta _{L}^{ +}  + \frac{{\pi}
}{{6}}} \right)^{2} + \mathrm{cos}\left( {\Delta _{L}^{ -} -
\frac{{\pi} }{{6}}} \right)^{2}} \right]\times &&\\\nonumber
\times \mathrm{ln}\left[ {\frac{{1}}{{1 - \mathrm{exp}\left( { -
\frac{{157890}}{{Tn_{eff}^{2}} }} \right)}}} \right]\quad
\,\,\left[ {\mathrm{s}^{ - 1}} \right].&&
\end{eqnarray}

\noindent Here, $A_{L} \sim 1$ is the new scaling coefficient,
which is only slightly different for \textit{nS}, \textit{nP} and
\textit{nD} Rydberg states of various alkali-metal atoms, in
contrast to \textit{C}$_{L}$ that ranges from 0.003 for lithium
\textit{nS} states to 1 for sodium \textit{nD} states. For
\textit{nS} states, only the first term in the square brackets of
equation~(\ref{eq27}) corresponding to
transitions with ${L}' = L + 1$ must be considered.

For estimates of the direct BBR photoionization rates with the
accuracy or 50\% it is sufficient to choose $A_{L} = 1$. For more
accurate calculations the values of \textit{A}$_{L}$ can be taken
from Table 2. We have obtained these values by comparing
the results of analytical and more accurate numerical
calculations. The coefficients are close to unity, except
\textit{nP} states of potassium and \textit{nD} states of rubidium
and cesium.

\begin{longtable}
{|p{75pt}|p{90pt}|p{90pt}|p{90pt}|} a & a & a & a  \kill \hline
\multicolumn{4}{|p{325pt}|}{\textbf{Table 2}. Numerically
determined scaling coefficients $A_{L}$ in equation~(\ref{eq27}).}
\\ \hline & \textit{A}$_{S}$\textit{}& \textit{A}$_{P}$\textit{}&
\textit{A}$_{D}$\textit{} \\ \hline Li& 1& 1& 0.9 \\ \hline Na& 1&
1& 1.1 \\ \hline K& 0.9& 0.45& 1.3 \\ \hline Rb& 1& 1& 0.6 \\
\hline Cs& 0.85& 1.1& 0.35 \\ \hline
\end{longtable}

\begin{figure}
\begin{center}
\includegraphics[width=13cm]{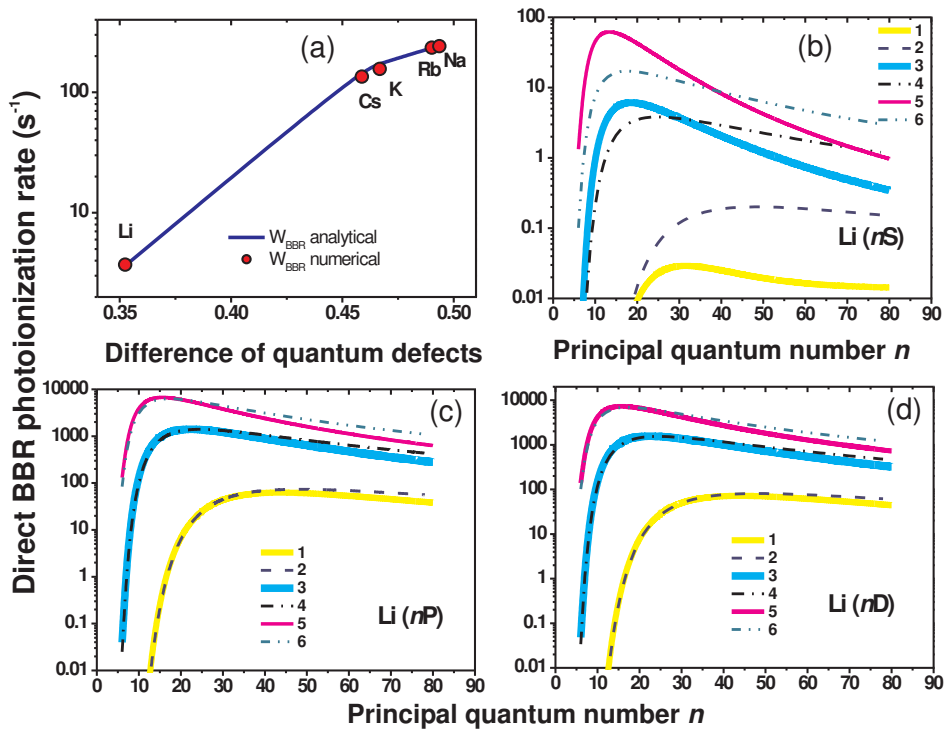}% Here is how to import EPS art
\caption{Direct BBR photoionization rates of alkali-metal Rydberg
states. (a)~Dependence of the photoionization rates of the
30\textit{S} state of lithium, sodium, potassium, rubidium, and
cesium at \textit{T}=300~K on the difference of quantum defects
$\mu_{S} - \mu_{P} $; (b), (c), (d) - dependence of the
photoionization rates of lithium \textit{nS, nP}, and \textit{nD}
Rydberg states on the principal quantum number $n$. Curves (1),
(3), (5) are numerical results obtained using the Dyachkov and Pankratov
model at the ambient temperatures \textit{T}=77, 300, and 600~K,
respectively. Curves (2), (4), (6) are analytical results obtained using
equation~(\ref{eq27}) at the ambient temperatures \textit{T}=77,
300, and 600~K, respectively.}
\label{Fig3}
\end{center}
\end{figure}

The results of our numerical and analytical calculations of the
direct BBR photoionization rates of alkali-metal Rydberg atoms are
presented in figure~\ref{Fig3}. Figure~\ref{Fig3}(a) shows the
dependence of \textit{W}$_{BBR}$ on the quantum defect for the
30\textit{S} Rydberg state of different alkali-metal atoms at the
temperature \textit{T}=300~K. A good agreement between the numerical
results and formula~(\ref{eq27}) with
\textit{A}$_{L}$=1 is found. For \textit{nP} and \textit{nD}
states such a simple dependence cannot be obtained because, in contrast to \textit{nS} states, there are two
ionization channels with ${L}' = L + 1$ and ${L}' = L - 1$
involved.
Figures~\ref{Fig3}(b)-(d) present the results for \textit{nS},
\textit{nP} and \textit{nD} states of lithium with \textit{n}=5-80
at three ambient temperatures \textit{T}=77, 300, and 600~K
(coefficients \textit{A}$_{L}$ are taken from Table 2). For
\textit{nP} and \textit{nD} states [figures~\ref{Fig3}(c) and (d)]
the formula~(\ref{eq27}) agrees well with the numerical results, while
for \textit{nS} states [figure~\ref{Fig3}(b)] the shapes of
numerical and analytical curves are completely different. This is
caused by strongly non-hydrogenic character of lithium
\textit{nS} states, which will be discussed in detail in
Section~2.3.

\begin{figure}
\begin{center}
\includegraphics[width=17cm]{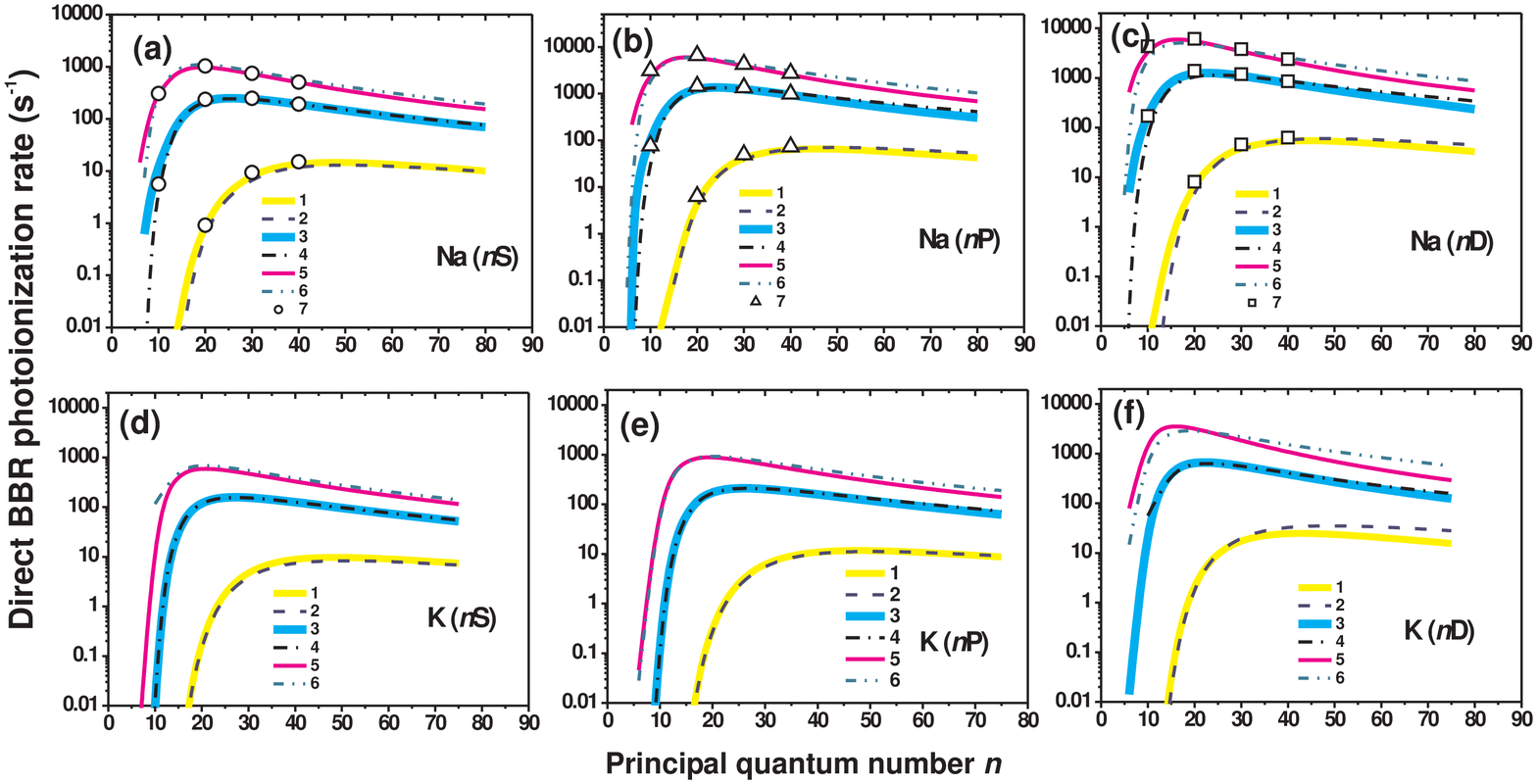}% Here is how to import EPS art
\caption{\label{Fig4} Direct BBR photoionization rates of
(a)~\textit{nS}, (b)~\textit{nP}, and (c)~\textit{nD} Rydberg states of sodium,
and (d)~\textit{nS}, (e)~\textit{nP}, and (f)~\textit{nD}
Rydberg states of potassium. Curves (1), (3), (5)
are numerical results obtained using the Dyachkov and Pankratov model at the
ambient temperatures \textit{T}=77, 300, and 600~K, respectively.
Curves (2), (4), (6) are analytical results obtained using
equation~(\ref{eq27}) at the ambient temperatures \textit{T}=77,
300, and 600~K, respectively. Curve (7) shows the numerical results for sodium
published by Lehman \cite{Lehman}.}
\end{center}
\end{figure}

\begin{figure}
\begin{center}
\includegraphics[width=17cm]{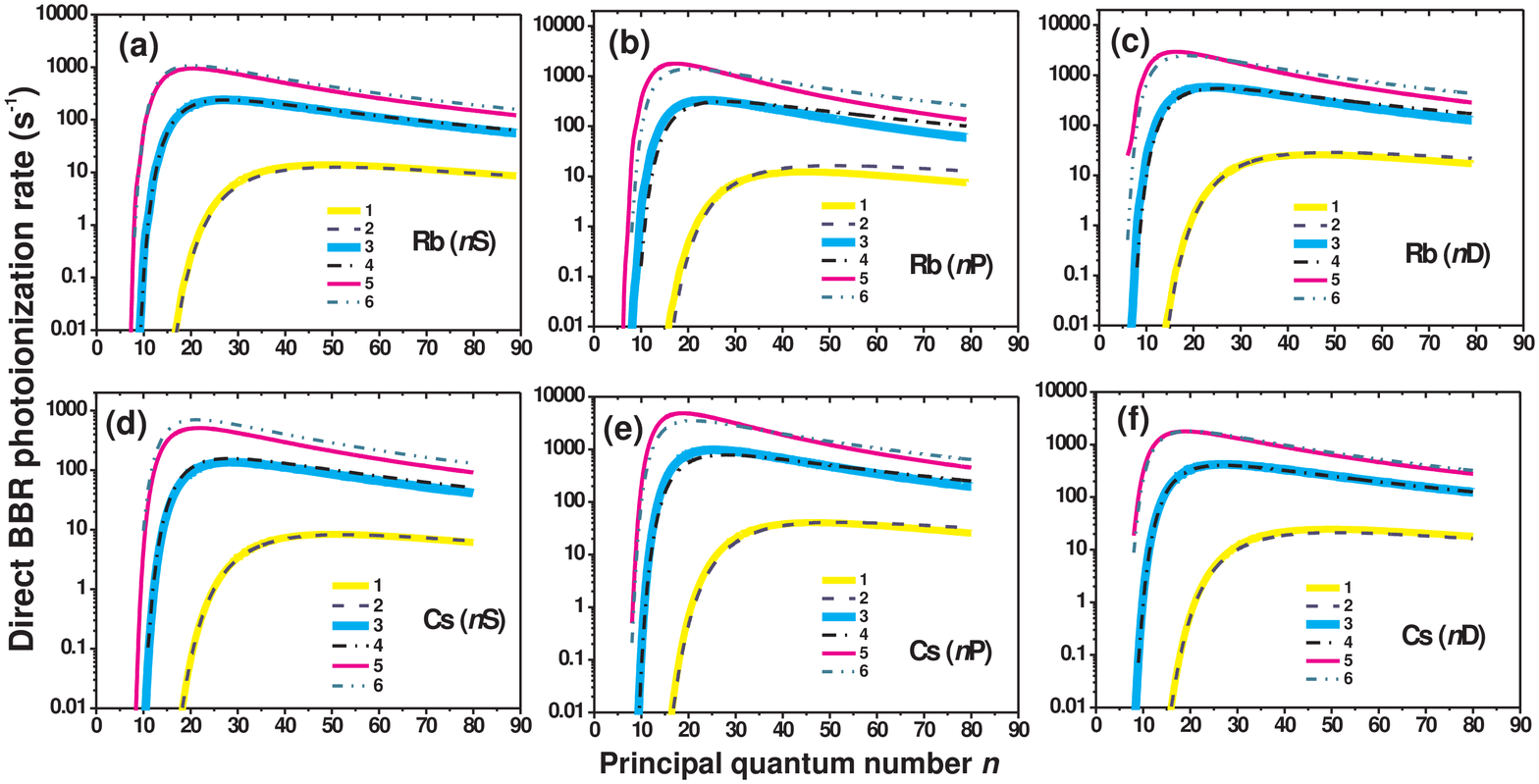}% Here is how to import EPS art
\caption{\label{Fig5} Direct BBR photoionization rates of
(a)~\textit{nS}, (b)~\textit{nP}, and (c)~\textit{nD} Rydberg states of rubidium,
and (d)~\textit{nS}, (e)~\textit{nP}, and (f)~\textit{nD}
Rydberg states of cesium. Curves (1), (3), (5)
are numerical results obtained using the Dyachkov and Pankratov model at the
ambient temperatures \textit{T}=77, 300, and 600~K, respectively.
Curves (2), (4), (6) are analytical results obtained using
equation~(\ref{eq27}) at the ambient temperatures \textit{T}=77,
300, and 600~K, respectively.}
\end{center}
\end{figure}

Results of calculations of \textit{W}$_{BBR}$ for Na, K, Rb, and Cs
atoms in \textit{nS}, \textit{nP}, and \textit{nD} Rydberg states
with \textit{n}=5-80 at the ambient temperatures of \textit{T}=77,
300, and 600~K are presented in figures~\ref{Fig4} and \ref{Fig5}.
In addition, our numerical and analytical calculations for Na are
compared with the results by Lehman \cite{Lehman} and a good
agreement is observed [see figure~\ref{Fig4}(a),(b),(c)]. For
other alkali-metal Rydberg states such a comparison is not
possible because, to the
best of our knowledge, no other published data are available.

A good agreement between the numerical and analytical results is
found for $n<50$. For higher \textit{n} the accuracy of the
analytical formula decreases and becomes worse than 50\% at
\textit{n}$\sim $100 for \textit{n}P states. For higher
\textit{L}, a neglected contribution from the terms proportional
to \textit{L} and \textit{L}$^{2}$ in equation~(\ref{eq27})
becomes more important. However, keeping these terms in
equation~(\ref{eq27}) complicates the formula but does not
substantially improve the accuracy.
%===============================================================================================
%=======================*************FIGURE 3**********=========================================
%===============================================================================================

%==============================================================================================

%===============================================================================================
%=======================*************FIGURE 4**********=========================================
%===============================================================================================

%==============================================================================================
%=======================*************FIGURE 5**********=========================================
%===============================================================================================

%==============================================================================================

We note that analytical formula~(\ref{eq27}) uses an asymptotic
expansion of the MacDonald functions, which is valid at $\omega
L^{3} < < 1$. In the slow-varying part of the
integral~(\ref{eq22}) in the square brackets we replaced $\omega $
by ${{1} \mathord{\left/ {\vphantom {{1} {2n_{eff}^{2}} }} \right.
\kern-\nulldelimiterspace} {2n_{eff}^{2}} }$. Such replacement
formally requires $\omega_{nL} > kT$ (at temperature
\textit{T}=300~K it is correct only for states with $n<20$).
Nevertheless, a comparison with our numerical results has shown
that equation~(\ref{eq27}) actually gives correct estimates of BBR
ionization rates also for higher values of \textit{n} (up to
\textit{n}$\sim $50). We conclude that equation~(\ref{eq27}) is
applicable for $L < < n$ and provides accurate estimates of
BBR-induced photoionization rates of \textit{nS}, \textit{nP} and
\textit{nD} alkali-metal Rydberg states.

\subsection{BBR-induced mixing of Rydberg states}

BBR causes not only direct photoionization of the initially
populated Rydberg levels. It also induces transitions between
neighboring Rydberg states, thus leading to population
redistribution \cite{Galvez1995, Galvez1997, PaperII}. For
example, after laser excitation of the Na $16S$ state, the
BBR-induced transitions populate the neighboring $n'P$ states
[figure~\ref{Fig1}(a)]. The calculations show that these states
have significantly higher direct photoionization rates than the
$16S$ state itself. Hence, BBR-induced population transfer to
$n'P$ states can noticeably affect the effective BBR
photoionization rate. The rates of spontaneous and BBR-induced
transitions from the initial $16S$ and $16D$ states to $n'P$
states have been calculated by us in \cite{PaperI} and are shown
in figures~\ref{Fig1}(b) and (c).

Importantly, absorption of BBR induces also transitions to higher
Rydberg states, which are denoted as $n''$ in
figure~\ref{Fig1}(a). These states can be ionized by the electric
field pulses usually applied in experiments in order to extract
ions into ionization detectors.

\subsection{Field ionization of high Rydberg states populated by
BBR}

Extraction electric-field pulses, which are commonly used to
extract ions from the ionization zone to the ionization detector,
ionize all Rydberg states with effective quantum numbers $n_{eff}$
exceeding some critical value $n_c$. This critical value $n_c$
depends on the amplitude of the applied electric field and it can
be found from the approximate formula~(\ref{eq9}). Hence, if a BBR
mediated process populates a state with $n' \ge n_c$, this state
will be ionized and thus will contribute to the detected
ionization signal \cite{SpencerIon1982}.

In order to analyze the efficiency of this process, we calculated
the radial matrix elements $R \left( nL \to n'L' \right)$ of
dipole-allowed transitions to other $n'L'$ states with $L'=(L \pm
1)$ using the semi-classical formulas of \cite{DyachkovPankratov}.
The total rate $W_{SFI}$ of BBR transitions to all Rydberg
states with $n' \ge n_c$ can be calculated by summing the
individual contributions of $nL \to n'L'$ transitions given by
equation~(\ref{eq6}):

\begin{equation}
\label{eq28} W_{SFI} = \,\sum\limits_{n' \ge n_{c}}
{\,\sum\limits_{L' = L \pm 1} {W\left( {nL \to n'L'} \right)}}.
\end{equation}

\noindent We have numerically calculated the values of $W_{SFI}$
for various amplitudes \textit{E} of the electric-field pulses.

We also compared the numerical values with those obtained from the
approximate analytical formulae, which have been derived using the bound-bound
matrix elements of the GDK model:

\begin{equation}
\label{eq29} W_{SFI} \approx \frac{{1}}{{\pi
^{2}c^{3}n^{3}}}\int\limits_{\frac{{1}}{{2n^{2}}} -
\frac{{1}}{{2n_{c}^{2} }}}^{\omega _{nL}}  {\left[ {2.22\,\omega
^{ - 1/3} + 2.63\,\omega ^{1/3}L^{2}} \right]} \frac{{d\omega}
}{{\mathrm{exp}\left(\omega/kT\right) - 1}}.
\end{equation}

\noindent The integration limits are chosen such that the integral
accounts for transitions to those Rydberg states for which
$\left( {\frac{{1}}{{2n^{2}}} - \frac{{1}}{{2n_{c}^{2}} }} \right)
< \omega < \omega _{nL} $ (i.e., states above the field ionization
threshold). Integration of equation~(\ref{eq29}) in the same
approximation as for equation~(\ref{eq22}) gives another useful
analytical formula that is similar to equation~(\ref{eq27}):

\begin{equation}
\label{eq30}
\begin{array}{l}
 W_{SFI} = A_{L} \frac{{11500T}}{{n^{7/3}}}\left[ {\mathrm{cos}\left( {\Delta_{L}^{
+}  + \frac{{\pi} }{{6}}} \right)^{2} + \mathrm{cos}\left( {\Delta
_{L}^{ -}  - \frac{{\pi} }{{6}}} \right)^{2}} \right] \times \\
 \times \;\left[ {\mathrm{ln}\frac{{1}}{{1 - \mathrm{exp}\left( {\frac{{157890}}{{Tn_{c}^{2}
}} - \frac{{157890}}{{Tn^{2}}}} \right)}} -
\mathrm{ln}\frac{{1}}{{1 - \mathrm{exp}\left( { -
\frac{{157890}}{{Tn^{2}}}} \right)}}} \right]\quad \mathrm{s}^{ -
1}
\\
 \\
 \end{array},
\end{equation}

\noindent
where \textit{T} is in Kelvins.

\begin{figure}
\begin{center}
\includegraphics[width=11cm]{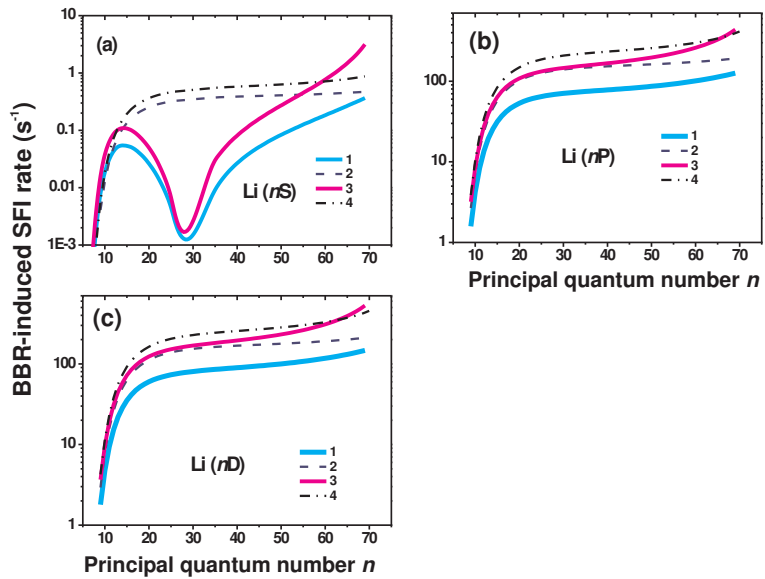}% Here is how to import EPS art
\caption{\label{Fig6} Calculated BBR-induced SFI rates
$W_{SFI}$ for (a)~\textit{nS}, (b)~\textit{nP}, and (c)~\textit{nD}
Rydberg states of lithium at the ambient temperature of 300~K.
Curves (1) and (3) are numerical results obtained using the Dyachkov and Pankratov
model at the electric-field amplitudes of $E$=5 and 10~V/cm,
respectively. Curves (2) and (4) are analytical results obtained using
equation~(\ref{eq30}) at the electric-field amplitudes of $E$=5
and 10~V/cm, respectively.}
\end{center}
\end{figure}

\begin{figure}
\begin{center}
\includegraphics[width=17cm]{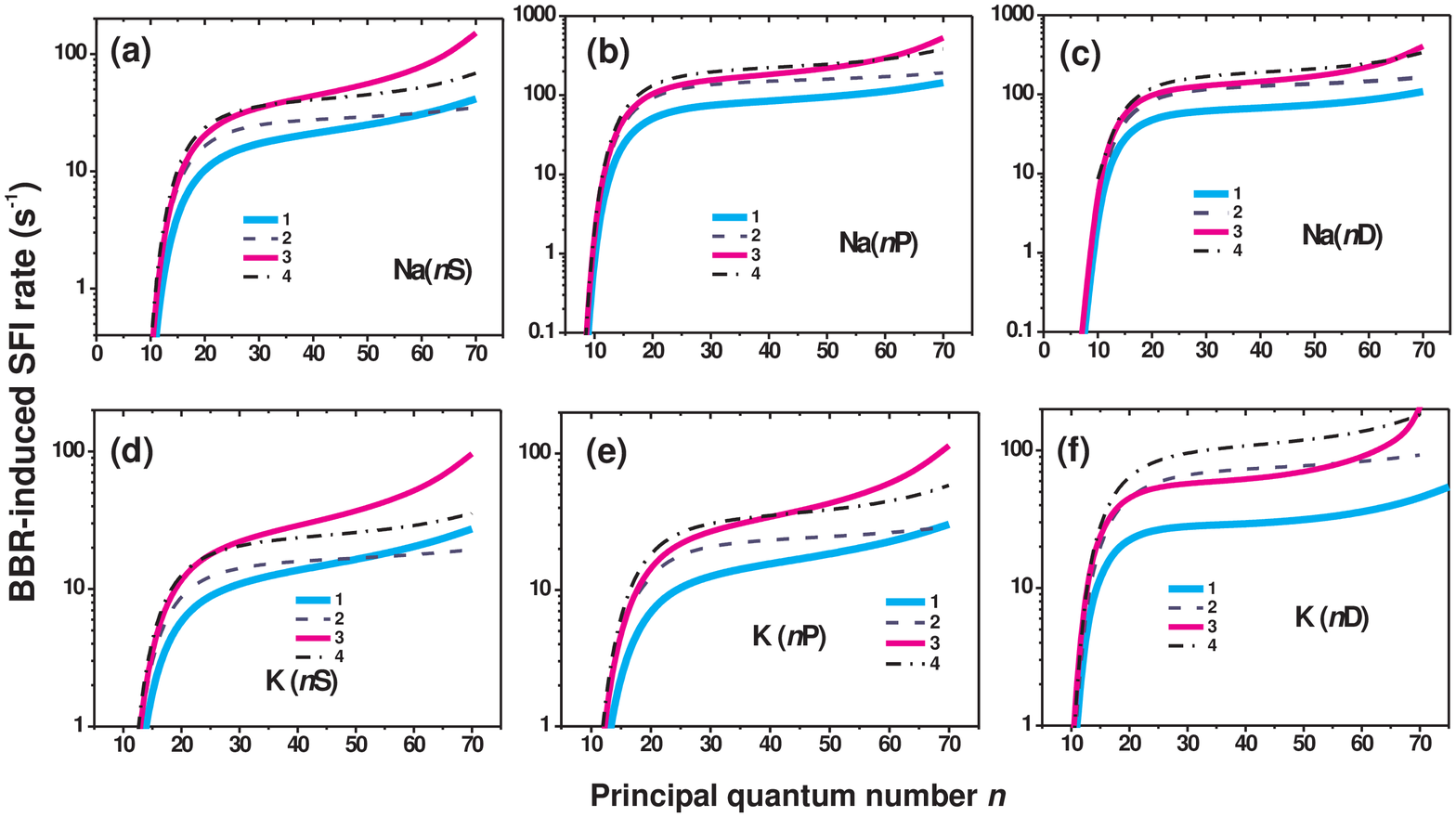}% Here is how to import EPS art
\caption{\label{Fig7} Calculated BBR-induced SFI rates
$W_{SFI}$ for
(a)~\textit{nS}, (b)~\textit{nP}, and (c)~\textit{nD} Rydberg states of sodium,
and (d)~\textit{nS}, (e)~\textit{nP}, and (f)~\textit{nD}
Rydberg states of potassium
at the ambient temperature of 300~K.
Curves (1) and (3) are numerical results obtained using the Dyachkov and Pankratov
model at the electric-field amplitudes of $E$=5 and 10~V/cm,
respectively. Curves (2) and (4) are analytical results obtained using
equation~(\ref{eq30}) at the electric-field amplitudes of $E$=5
and 10~V/cm, respectively.}
\end{center}
\end{figure}

\begin{figure}
\begin{center}
\includegraphics[width=17cm]{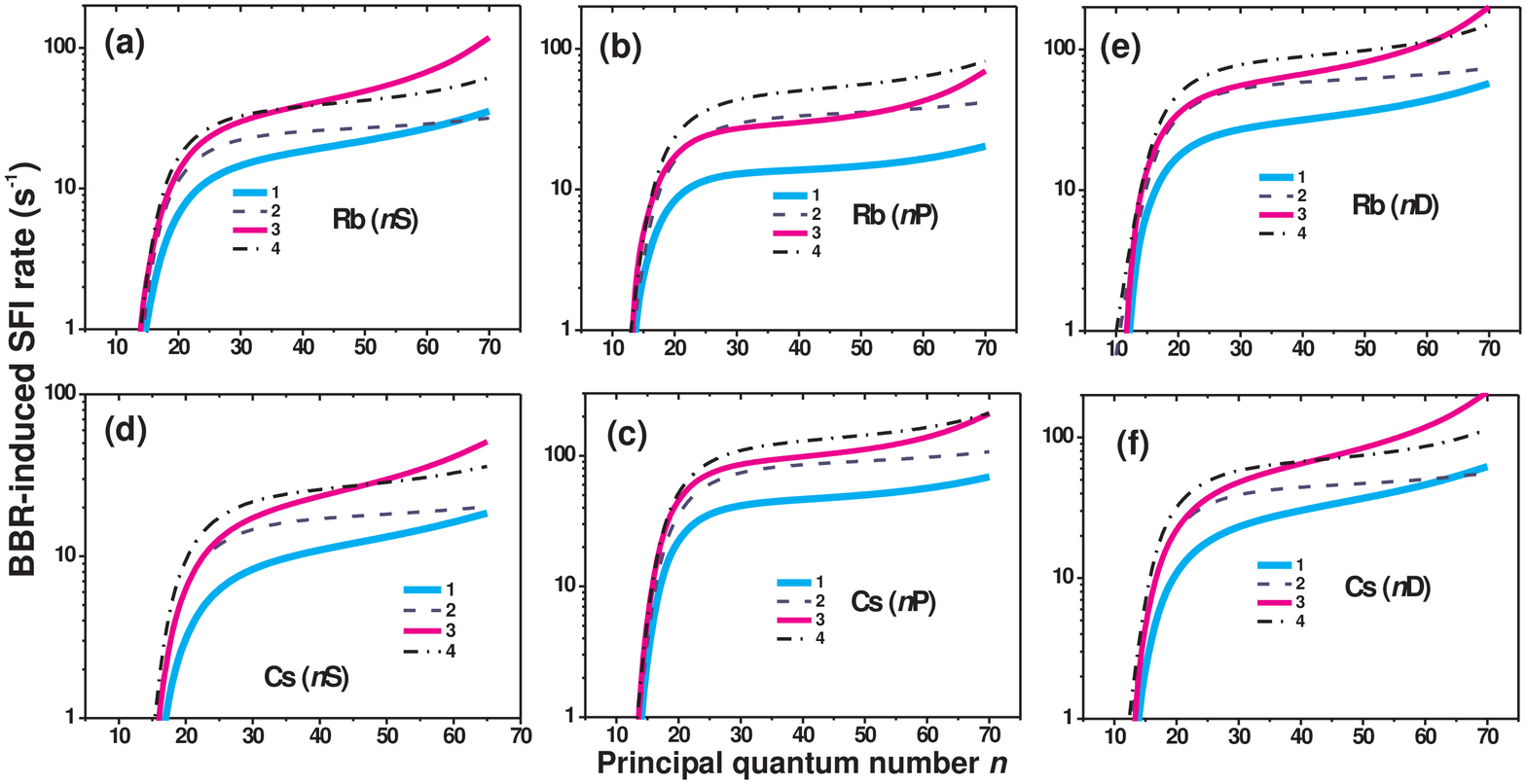}% Here is how to import EPS art
\caption{\label{Fig8} Calculated BBR-induced SFI rates
W$_{SFI}$ for
(a)~\textit{nS}, (b)~\textit{nP}, and (c)~\textit{nD} Rydberg states of rubidium,
and (d)~\textit{nS}, (e)~\textit{nP}, and (f)~\textit{nD}
Rydberg states of caesium
at the ambient temperature of 300~K.
Curves (1) and (3) are numerical results obtained using the Dyachkov and Pankratov
model at the electric-field amplitudes of \textit{E}=5 and
10~V/cm, respectively. Curves (2) and (4) are analytical results
obtained using equation~(\ref{eq30}) at the electric-field amplitudes of
\textit{E}=5 and 10~V/cm, respectively.}
\end{center}
\end{figure}

The obtained numerical and analytical data on $W_{SFI}$ calculated
for \textit{nS}, \textit{nP}, and \textit{nD} alkali-metal Rydberg
states with \textit{n}=5-80 at the ambient temperatures
\textit{T}=77, 300, and 600~K and the amplitudes of the electric
field of 5 and 10~V/cm are presented in
figures~\ref{Fig6}-\ref{Fig8}. The scaling coefficients $A_{L}$
from Table~2 have been used.

We have unexpectedly found that the dependence of $W_{SFI}$ on
\textit{n} for lithium \textit{nS} Rydberg states exhibits a deep
minimum at $n\sim $30 [figure~\ref{Fig6}(a)]. For \textit{nP} and
\textit{nD} states of lithium and \textit{nS}, \textit{nP} and
\textit{nD} states of other alkali-metal Rydberg atoms such a
minimum is absent [figures~\ref{Fig7}-\ref{Fig8}]. A theoretical
analysis has shown that this anomaly is caused by a Cooper minimum
in the discrete spectrum for transitions between \textit{nS} and
$n'P$ lithium Rydberg states \cite{CooperLi}. It can be explained
as follows. For hydrogen atoms the radial matrix elements of
transitions between bound Rydberg states decrease monotonously
with increase of the interval between energy levels. In contrast,
the radial wave functions of alkali-metal Rydberg states have
varying phase shifts $\pi \mu_l$, which can suppress the
overlapping between wave functions in the calculated radial matrix
elements of the electric dipole moment \cite{FanoCooper}. This
leads to a minimum in transition probabilities, which are
proportional to the square of radial matrix elements. The
hydrogenic GDK model [dashed curve in figure~\ref{Fig6}(a)] does
not predict a Cooper minimum, since the phase shifts of radial
wave functions due to quantum defects are ignored.

For hydrogen-like \textit{nP} and \textit{nD} lithium Rydberg
states [figure~\ref{Fig6}(b) and (c), respectively] the analytical
and numerical calculations give close results, although the
agreement between them is worse than in the case of the direct photoionization
by BBR. The same is observed for
\textit{nS},\textit{ nP} and \textit{nD} Rydberg states of sodium,
potassium, rubidium, and cesium [figures~\ref{Fig7}-\ref{Fig8}].

The chosen amplitudes of the electric field of 5 and 10~V/cm
correspond to the typical conditions of experiments with
laser-cooled Rydberg atoms, because such fields do not lead to
ionization of Rydberg states with large quantum numbers $n\sim
$30-50 relevant to experiments on ultracold plasma
\cite{LiNoele2004}. Such electric fields are just sufficient to extract from the interaction volume
ions formed due to collisional and BBR-induced ionization of
Rydberg atoms. At
$n \sim $30 the rate of ionization by electric field is by an order
of magnitude smaller than the rate of direct BBR-induced
photoionization.
The rates of direct BBR photoionization and BBR-induced SFI become
comparable at \textit{n}$\sim $60.

\begin{figure}
\begin{center}
\includegraphics[width=13cm]{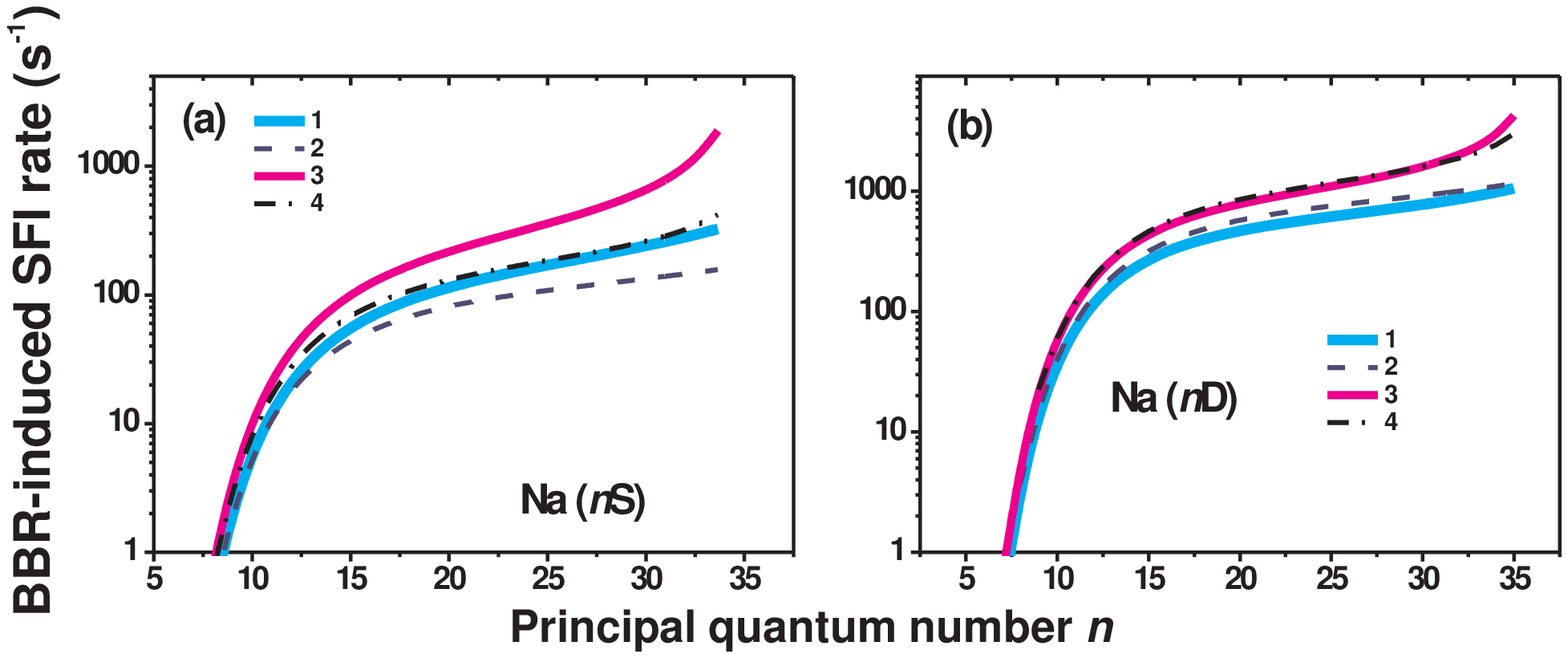}% Here is how to import EPS art
\caption{\label{Fig9} Calculated BBR-induced SFI rates
$W_{SFI}$ for (a)~\textit{nS} and (b)~\textit{nD} Rydberg
states of sodium at the ambient temperature of 300~K. Curves (1) and (3)
are numerical results obtained using the Dyachkov and Pankratov model at the
electric-field amplitudes of \textit{E}=100 and 200~V/cm,
respectively. Curves (2) and (4) are analytical results obtained using
equation~(\ref{eq30}) at the electric-field amplitudes of
\textit{E}=100 and 200~V/cm, respectively.}
\end{center}
\end{figure}

In our experiment (subection~3.1) we used sodium Rydberg atoms
in states with low $n\sim $8-20, which were interesting in the context of
collisional ionization studies. The experimental conditions required the
use of extracting electric pulses of larger amplitude
(100~V/cm). The results of the calculation of $W_{SFI}$ for
\textit{nS} and \textit{nD} sodium Rydberg atoms by the 100~V/cm
and 200~V/cm electric pulses are shown in figure~\ref{Fig9}. The
calculations were made only for \textit{n}=5-35, since Rydberg
states with $n>37$ are ionized by the 200~V/cm electric field. For
$n\sim $20 the rate of direct BBR photoionization is
only two times larger than $W_{SFI}$. Hence, the contribution of BBR-induced SFI is important.

%==============================================================================================
%=======================*************FIGURE 6**********=========================================
%===============================================================================================

%==============================================================================================

%==============================================================================================
%=======================*************FIGURE 7**********=========================================
%===============================================================================================

%==============================================================================================

%==============================================================================================
%=======================*************FIGURE 8**********=========================================
%===============================================================================================

%=========================================================================
%=========================================================================
%=========================================================================
\subsection{Total BBR-induced ionization rates}
%=========================================================================
%=========================================================================

\begin{figure}
\begin{center}
\includegraphics[width=7cm]{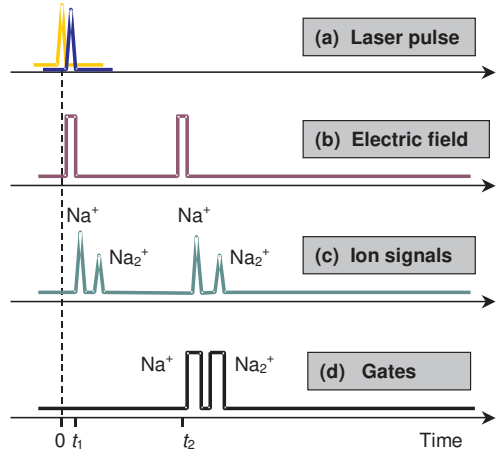}% Here is how to import EPS art
\caption{\label{Fig10} Timing diagram of the experiment: (a) laser
excitation pulse; (b) electric field pulses for the ion
extraction; (c) atomic $A^{+}$ and molecular $A_{2}^{+}$ ion
signals; (d) detector gates.}
\end{center}
\end{figure}

In this section we shall analyze the time evolution of populations
of Rydberg states during the interaction with ambient BBR photons.
A typical timing diagram for the laser excitation of Rydberg
states and detection of ions created by SFI is shown in figure~\ref{Fig10}.
This scheme was used in our recent experiment on collisional
ionization of Na Rydberg atoms \cite{PaperI}. The electric field
was formed by two metallic plates, one of them having a hole
with a mesh allowing the extraction of ions. Two identical electric field
pulses with the 100~V/cm amplitude and 250~ns duration
[figure~\ref{Fig10}(b)] were applied to the repelling plate after
each laser excitation pulse [figure~\ref{Fig10}(a)]. The first
pulse was applied immediately after the laser pulse to remove the
atomic $A^{+}$ and the molecular $A_2^+$ ions produced during the
laser pulse. The second pulse extracted to a particle detector
(channeltron) those ions, which appeared in the time interval
between $t_1=0.3 \, \mu$s and $t_2=2.1 \, \mu$s after the laser
excitation pulse. These ions appeared due to collisional and BBR-induced
ionization of Rydberg atoms. In the mass spectrum detected by the
channeltron, the signals of the atomic A$^{+}$ and the molecular
A$_{2}^{+}$ ions were separated by 0.6~$\mu$s and thus well
resolved [figure~\ref{Fig10}(c)]. The gated pulse counters
registered the signals from the atomic and molecular ions
independently [figure~\ref{Fig10}(d)].

%==============================================================================================
%==============================================================================================
%=======================*************FIGURE 9**********=========================================
%===============================================================================================

%==============================================================================================

Let us first consider the simplest case of laser excitation of a
single sodium \textit{nS} state. The evolution of the number
$N_{A^+}$ of atomic ions produced via absorption of BBR photons by
atoms in the initial \textit{nS} state is given by

\begin{equation}
\label{eq31}
\frac{{dN_{A^{ +} } \left( {t} \right)}}{{dt}} = W_{BBR} N_{nS} \left( {t}
\right),
\end{equation}

\noindent where $N_{nS} \left( {t} \right) = N_{nS}
\left( {t = 0} \right)\mathrm{exp}\left( { - t/\tau _{eff}^{nS}}
\right)$ is the total number of Rydberg atoms remaining in the
\textit{n}S state as a function of time, and $\tau _{eff}^{nS} $
is the effective lifetime of the \textit{nS} state. The number of photoions produced by BBR during
the time interval $(t_1, t_2)$ is of interest. The total number of ions produced
during this interval by direct BBR photoionization of the
\textit{nS} state can be found by integrating
equation~(\ref{eq31}) from $t_1$ to $t_2$:

\begin{equation}
\label{eq32} N_{A^{ +} } = N_{nS} \left( {t = 0} \right)W_{BBR}
\tau _{eff}^{nS} \left[ {\mathrm{exp}\left( { - t_{1} /\tau
_{eff}^{nS}} \right) - \mathrm{exp}\left( { - t_{2} /\tau
_{eff}^{nS}} \right)} \right].
\end{equation}

\noindent This result can be rewritten by introducing an effective
interaction time $t_{eff}^{nS} $ \cite{PaperI}:

\begin{eqnarray}
\label{eq33} N_{A^{ +} } = N_{nS} \left( {t = 0} \right)W_{BBR}
t_{eff}^{nS} , &&\\\nonumber t_{eff}^{nS} = \tau _{eff}^{nS}
\left[ {\mathrm{exp}\left( { - t_{1} /\tau _{eff}^{nS}}  \right) -
\mathrm{exp}\left( { - t_{2} /\tau _{eff}^{nS}}  \right)}
\right].&&
\end{eqnarray}

Blackbody radiation induces also
transitions to other Rydberg states $n'P$, as discussed in
section~2.2. Evolution of populations of these states is
described by the rate equation

\begin{eqnarray}
\label{eq34} \frac{{dN_{{n}'P} \left( {t} \right)}}{{dt}} = \left[
{W\left( {nS \to {n}'P} \right) + A\left( {nS \to {n}'P} \right)}
\right]N_{nS} \left( {t} \right)-&& \\\nonumber \qquad \qquad
\qquad - {{N_{{n}'P} \left( {t} \right)} \mathord{\left/
{\vphantom {{N_{{n}'P} \left( {t} \right)} {\tau _{eff}^{n'P}} }}
\right. \kern-\nulldelimiterspace} {\tau _{eff}^{n'P}} },&&
\end{eqnarray}

\noindent where $A(nS \to n'P)$ and $W(nS \to n'P)$ are the rates
of population of $n'P$ states due to spontaneous transitions and BBR-induced transitions from the
initial \textit{nS} state, respectively, and $\tau _{eff}^{n'P} $
is the effective lifetime of the $n'P$ state.

A combination of equation~(\ref{eq34}) with equations~(\ref{eq31})
and (\ref{eq32}) yields

\begin{eqnarray}
\label{eq35} W_{BBR}^{mix} \left( {nS} \right) =
\sum\limits_{{n}'} {\frac{{\left[ {W\left( {nS \to {n}'P} \right)
+ A\left( {nS \to {n}'P} \right)} \right]\,W_{BBR} \left( {{n}'P}
\right)}}{{{{1} \mathord{\left/ {\vphantom {{1} {\tau
_{eff}^{{n}'P}} }} \right. \kern-\nulldelimiterspace} {\tau
_{eff}^{{n}'P}} } - {{1} \mathord{\left/ {\vphantom {{1} {\tau
_{eff}^{nS} }}} \right. \kern-\nulldelimiterspace} {\tau
_{eff}^{nS}} }}}\,} \times &&\\\nonumber \qquad \qquad \qquad
\qquad \times \left( {1 - \frac{{t_{eff}^{nS}} }{{t_{eff}^{{n}'P}}
}} \right).
\end{eqnarray}

\noindent The main contribution to the sum in
equation~(\ref{eq35}) is from $n'P$ states with $n' =n \pm 1$ [see
figure~\ref{Fig1}(b)].

%==============================================================================================
%==============================================================================================
%=======================*************FIGURE 10**********=========================================
%===============================================================================================

%==============================================================================================

The effective BBR-induced ionization rates for \textit{nP} and \textit{nD}
states are determined in the same way as for \textit{nS} states,
taking into account the population transfer to both
$n'(\textit{L}+1)$ and $n'(\textit{L}-1)$ states.

The rate $W_{SFI}^{mix} $ describes the second-order process -
BBR-induced transitions from the neighboring $n'L'$ states to
highly excited states $n''L''$ with $n''>n_c$ [see
figure~\ref{Fig1}(a)], followed by ionization of these states by
extraction electric field pulses. This rate is also calculated
using equation~(\ref{eq35}), in which $W_{BBR}$ is replaced by
$W_{SFI}$ and the summation is done over the states with $n' <
n_c$.

\begin{figure}

\begin{center}
\includegraphics[width=17cm]{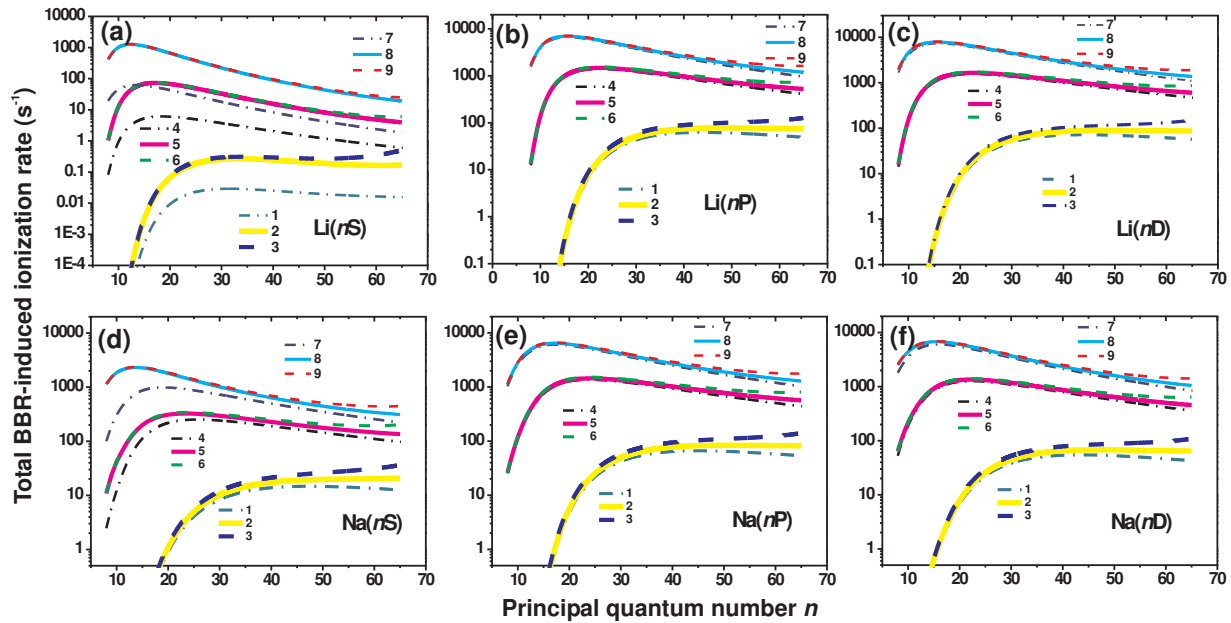}% Here is how to import EPS art
\caption{\label{Fig11} Calculated total BBR-induced ionization
rates $W_{BBR}^{tot}$ for (a)~\textit{nS}, (b)~\textit{nP},
(c)~\textit{nD} Rydberg states of lithium and (d)~\textit{nS},
(e)~\textit{nP}, and (d)~\textit{nD} Rydberg states of sodium.
Curves (1), (4), and (7) are the direct BBR-induced photoionization rates at the
ambient temperatures of \textit{T}=77, 300, and 600~K, respectively.
Curves (2), (5), and (8) are the total BBR-induced ionization rates at
the amplitude of extraction electric field pulses of $E$=5~V/cm
for ambient temperatures of \textit{T}=77, 300, and 600~K,
respectively. Curves (3), (6), and (9) are the total BBR-induced
ionization rates at the amplitude of extraction electric-field
pulses of $E$=10~V/cm for ambient temperatures of \textit{T}=77, 300, and
600~K respectively.}
\end{center}
\end{figure}

\begin{figure}

\begin{center}
\includegraphics[width=17cm]{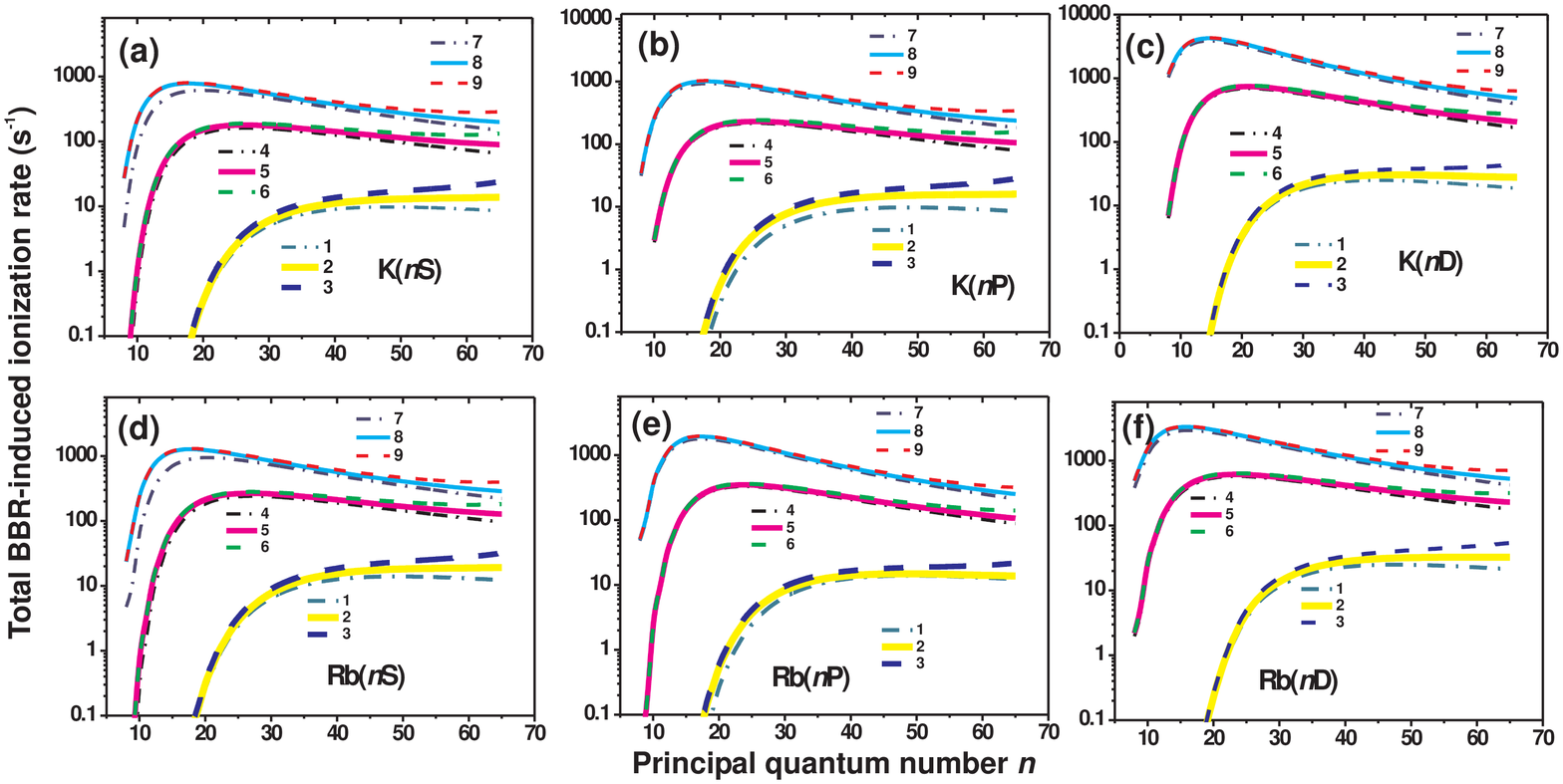}% Here is how to import EPS art
\caption{\label{Fig12}Calculated total BBR-induced ionization
rates $W_{BBR}^{tot} $ for (a)~\textit{nS}, (b)~\textit{nP}, and
(c)~\textit{nD} Rydberg states of potassium, and (d)~\textit{nS},
(e)~\textit{nP}, and (d) \textit{nD} Rydberg states of rubidium.
Curves (1), (4), (7) are the direct BBR-induced photoionization
rates for ambient temperatures of \textit{T}=77, 300, and 600~K,
respectively. Curves (2), (5), (8) are the total BBR-induced
ionization rates at the amplitude of extraction electric field
pulses of $E$=$5$~V/cm and ambient temperatures of \textit{T}=77,
300, and 600~K respectively. Curves (3), (6), (9) are the total
BBR-induced ionization rates at the amplitude of extraction
electric field pulses of $E$=10~V/cm and ambient temperatures of
\textit{T}=77, 300, and 600~K, respectively.}
\end{center}
\end{figure}

\begin{figure}

\begin{center}
\includegraphics[width=13cm]{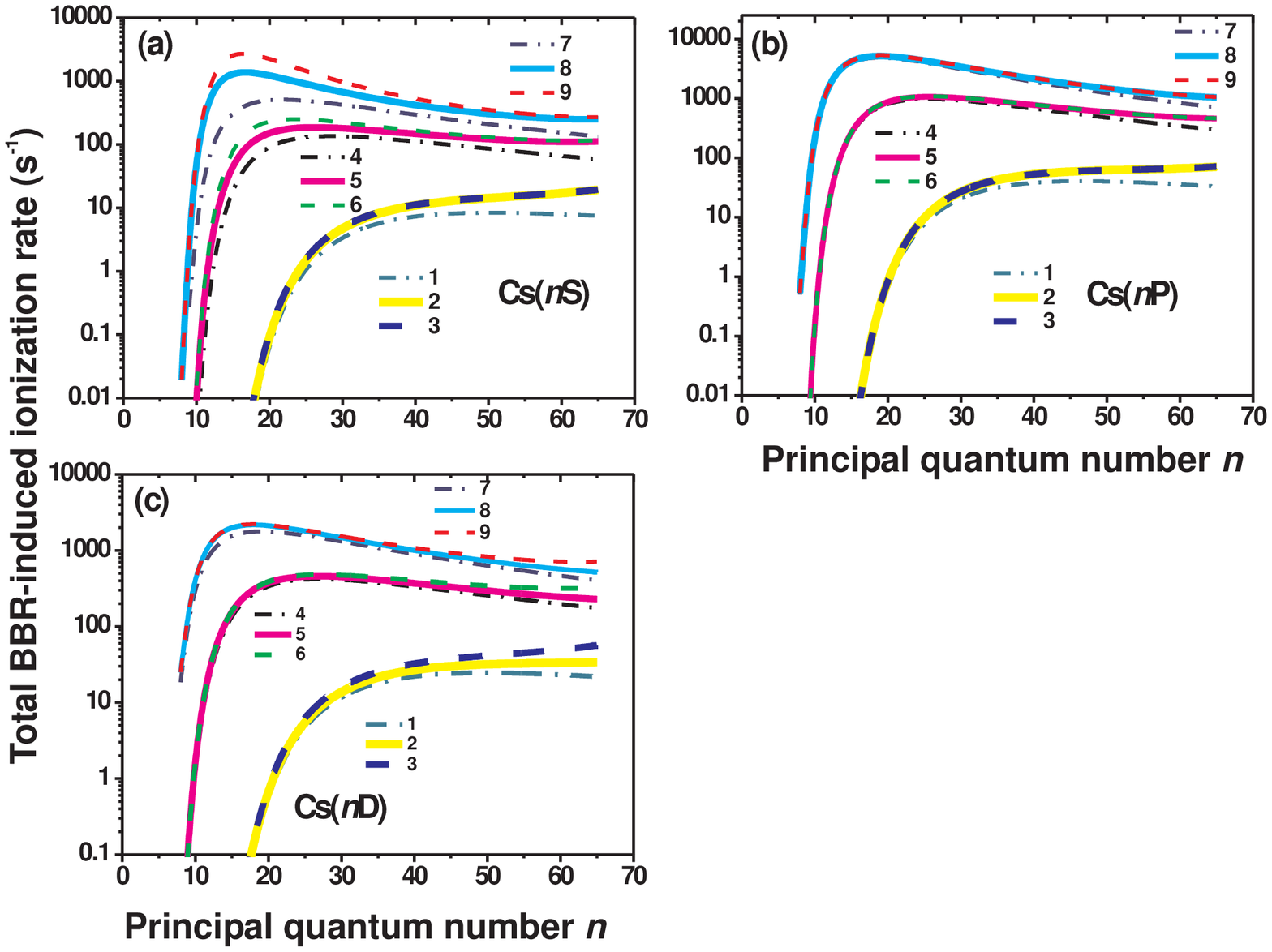}% Here is how to import EPS art
\caption{\label{Fig13}Calculated total BBR-induced ionization
rates $W_{BBR}^{tot} $ for (a)~\textit{nS}, (b)~\textit{nP}, and
(c)~\textit{nD} Rydberg states of cesium. Curves (1), (4), (7) are
the direct BBR-induced photoionization rates for ambient
temperatures of \textit{T}=77, 300, and 600~K, respectively.
Curves (2), (5), (8) are the total BBR-induced ionization rates at
the amplitude of extraction electric field pulses of $E$=5~V/cm
for ambient temperatures of \textit{T}=77, 300, and 600~K,
respectively. Curves (3), (6), (9) are the total BBR-induced
ionization rates at the amplitude of extraction electric-field
pulses of $E$=10~V/cm for ambient temperatures of \textit{T}=77,
300, and 600~K, respectively.}
\end{center}
\end{figure}

Figures~\ref{Fig11}-\ref{Fig13} show the results of numerical
calculations of total ionization rates $W_{BBR}^{tot}$ of
\textit{nS}, \textit{nP}, and \textit{nD} Rydberg states of alkali-metal
atoms. The calculations were done for a
broad range of principal quantum numbers (\textit{n}=8-65) and
temperatures (\textit{T} = 77, 300, 600~K), at the amplitudes of
extraction electric field of 5~V/cm (solid curves) and 10~V/cm
(dashed curves). For comparison, the direct BBR-induced ionization
rates are also shown (dash-dotted curves).

The values of $W_{BBR}^{tot}$ depend on the time interval of
accumulation of ions in the interaction region. For sodium and
rubidium atoms, the calculations were made for $t_1=0.3$~$\mu$s
and $t_2=2.1$~$\mu$s, which corresponds to the conditions of our
experiment described in Section~3. For lithium, potassium and
cesium atoms for the sake of simplicity we used $t_1=0$~$\mu$s
and $t_2=2$~$\mu$s. Such choice of the time interval does not
noticeably change the calculated rates; it is important
only for the states with low $n \sim 10$ with short lifetimes
$\sim 1$~$\mu$s. The effective lifetimes of Rydberg states,
which are necessary for the determination of $t_{eff}$ , were calculated using the
Dyachkov and Pankratov formulas for radial matrix elements
\cite{Lifetimes}.

Figure~\ref{Fig11}(a) shows the calculated
total BBR-induced ionization rates for lithium \textit{nS}
Rydberg states. Account for BBR-induced mixing leads to a
strong increase of the total BBR-induced ionization rate. In contrast, the rate
of the direct BBR-induced photoionization of lithium \textit{nP} Rydberg
states is by two orders of magnitude larger than the direct BBR-induced rate for \textit{nS} states, and the main
contribution to the total number of ions is due to the $n'P$ states with $n' = n
\pm 1$. Taking into account the SFI of high-lying Rydberg states and
photoionization of neighboring Rydberg states by BBR substantially
alters both the absolute values of $W_{BBR}^{tot}$ and the shapes
of their dependences on \textit{n}. In the case of
sodium, potassium, rubidium, and cesium Rydberg states
(figures~\ref{Fig11}-\ref{Fig13}) the difference between the
direct BBR-induced photoionization rates and the total BBR-induced
ionization rates is smaller, but remains observable.

%==============================================================================================
%==============================================================================================
%=======================*************FIGURE 11**********=========================================
%===============================================================================================

%==============================================================================================

%==============================================================================================
%==============================================================================================
%=======================*************FIGURE 12**********=========================================
%===============================================================================================

%==============================================================================================
%==============================================================================================
%==============================================================================================
%=======================*************FIGURE 13**********=========================================
%===============================================================================================

%==============================================================================================

%======================================================================
%======================================================================
%======================================================================
%======================================================================

\section{Experimental study of BBR-induced ionization}

\subsection{Temperature dependence of BBR-induced ionization rate}
%======================================================================
%======================================================================
%======================================================================
%======================================================================

The temperature dependence of BBR-induced ionization rate of the
sodium 17\textit{D} Rydberg atoms was measured by Spencer et al.
\cite{SpencerIon1982}. The measurements were performed in an
effusive beam of sodium atoms which passed through a thermally
controlled interaction region. The atoms were excited to the
17\textit{D} state by two pulsed dye lasers. The laser beams were
collinear with the atomic beam and excitation took place between
two electric-field plates in a cryogenic environment. After a
delay of 500~ns, a small electric field of 8~V/cm was applied to
guide the ions from the interaction region through a small hole in
the lower field plate to the electron multiplier. The 500~ns
interval was sufficiently short to ensure that BBR-induced
transitions to neighboring $n'P$ and $n'F$ levels did not affect
the results of measurements.

The temperature was varied in the range 90-300~K by pouring
measured amounts of liquid nitrogen into the cryostat and allowing
it to boil off. Although the cryostat rapidly equilibrated to the
new temperature, a thermal drift limited the useful observation
time to about 15 minutes for each data point, so that there were
about 1800 data pulses at each temperature. The number density of
excited atoms was less than $10^5 \,\mathrm{cm}^{-3}$, low enough
to avoid Rydberg-Rydberg collisions and superradiant transfer
between Rydberg states. The background pressure was less than
$10^{-7}$ Torr, sufficiently small to avoid also collisions with
background gases. The absolute collisional ionization rates were
not measured, but the experimental points were normalized to the
theory.

The three sources of possible uncertainties were considered by the
authors. The first one was a counting statistics, the other ones
were the fluctuations of the intensity of laser radiation,
responsible for errors of 2-3\% at each data point, and the last
one was a drift of the temperature of the vacuum chamber, which
led to error of $ \pm 5 \, \mathrm{s}^{-1}$ for each data point.

A systematic shift due to an extracting field was considered. The
8~V/cm field ionized all states with $n>80$ (see Section~2). The
calculated rate of BBR-induced transfer to states with $n>80$ was
so small that it could be neglected. However, an increase of the
amplitude to several hundred~V/cm, required to ionize the states
with $n>30$, led to a significant increase of the measured signal,
which was consistent with the results of the calculation of the
BBR-induced transfer rates to states with $n>30$.

BBR-induced ionization occurs due to photons of shorter wavelength
than those which cause transitions to neighboring Rydberg states.
Hence, a measurement of BBR-induced ionization rates instead of
discrete transition rates can be a stricter test of the
temperature of blackbody radiation. In ref.~\cite{SpencerIon1982},
the photoionization became observable at 100~K due to 1\%
parasitic contribution from 300~K blackbody radiation. Based on
the apparatus size and the emissivity of the materials surrounding
the interaction region, it was estimated that less than 0.4\% of a
300~K radiation existed within the interaction region. Finally,
the measured photoionization rate varied by a factor greater than
100 over the temperature range 77-300~K studied, and a good
agreement between experiment and theory was observed.

\subsection{Experimental study of the dependence of ionization rates on $n$.}

Burkhardt et al.~\cite{Burkhardt1986} studied the ionization of
sodium Rydberg atoms in a gas cell at the temperature of 500~K.
The atoms were excited to Rydberg \textit{nS} and \textit{nD}
states with $18 \le n \le 35$ by a pulsed dye laser. The
dependences of the ionization signals on the principal quantum
number were measured. It has been shown that at the number density
of ground-state atoms $\mathrm{n}_{3S}\sim 10^{11} \,
\mathrm{cm}^{-3}$ the photoionization by blackbody radiation was a
predominant source of atomic ions, and contribution from
collisional ionization could be neglected.

Allegrini et al.~\cite{AllegriniBa1988} studied collisional mixing
of Ba Rydberg states. The signal of BBR-induced ionization was
used to measure the relative population of barium Rydberg states,
instead of the commonly used SFI method. The number density of
ground-state atoms in the experiment was close to
$10^{12}\,\mathrm{cm}^{-3}$. At this density the method of SFI is
inapplicable due to electric breakdown. The simple
formula~(\ref{eq17}) was used to calculate the rate of BBR-induced
ionization.

\begin{figure}

\begin{center}
\includegraphics[width=10cm]{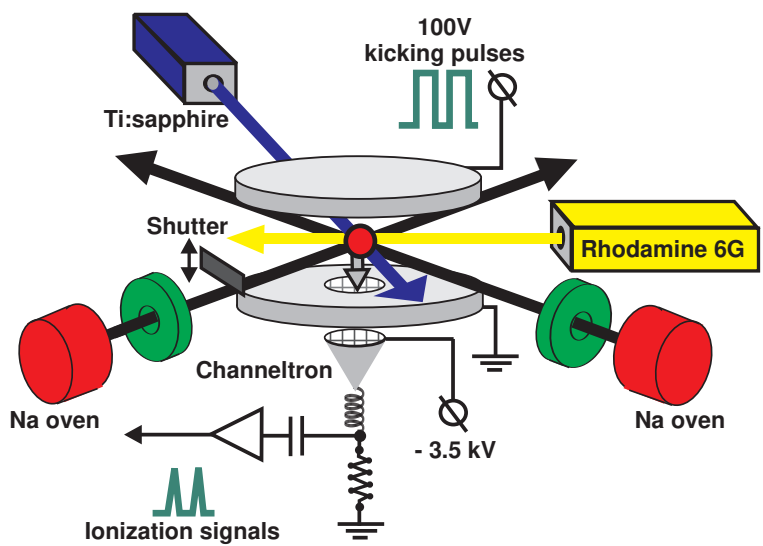}% Here is how to import EPS art
\caption{\label{Fig14} Experimental arrangement of the atomic and
laser beams, and the ion detection system.}
\end{center}
\end{figure}

The dependence of associative and BBR-induced ionization rates of
the sodium \textit{nS} and \textit{nD} Rydberg atoms with
\textit{n}=8-20 on the principal quantum number $n$ was measured
by us in~\cite{PaperI}. Experiments were performed using a single
effusive Na atomic beam in a vacuum chamber at the background
pressure of $5 \times 10^{-7}$~Torr (figure~\ref{Fig14}). The
temperature of the Na oven was stabilized at 635~K. The atomic
beam was formed by an expansion of sodium vapor through a 2~mm
diameter opening in the oven at a distance of 9~cm from the
reaction zone. Collimation of the beam was achieved by a 1.5~mm
diameter aperture, located 4~cm upstream from the reaction zone.
The effective diameter of the atomic beam in the reaction zone was
about 4~mm.

Sodium \textit{nS} and \textit{nD} Rydberg states were excited
using the two-step scheme $3S_{1/2} \to 3P_{3/2}  \to nS, nD$  by
radiations of two tunable lasers pulsed at a 5~kHz repetition
rate. In the first step, 50 ns pulses from a Rhodamine 6G
dye-laser with linewidth of 50 GHz were used. They saturated the
$3S_{1/2} \to 3P_{3/2}$ transition at 589~nm (\textit{yellow}).
The resonance fluorescence on this transition was detected by a
photomultiplier to monitor the relative changes in the number
density of the atomic beam. In the second step, the second
harmonic of a Ti-sapphire laser was used. It yielded 50~ns pulses
with 10~GHz linewidth, tunable in the 400-430~nm range
(\textit{blue}). When not focused, this radiation did not saturate
the $3P_{3/2} \to nS,nD$ transitions. The two laser beams were
crossed at a right angle in the reaction zone, both of them
crossing the atomic beam at a 45$^{\circ}$ angle. Laser beams were
spatially limited by 2~mm diameter apertures at the entrance
windows of the vacuum chamber. Such configuration ensured a
sufficiently small excitation volume of 2~mm size in the central
part of the atomic beam, where the spatial variation of atom
number density was insignificant $(<20\%)$.

%==============================================================================================
%==============================================================================================
%=======================*************FIGURE 14**********=========================================
%===============================================================================================

%==============================================================================================

The ion detection system shown in figure~\ref{Fig14} used a
channeltron multiplier VEU-6. The atomic beam passed between two
stainless-steel plates with diameter of 70~mm, spaced by 10~mm.
The plates formed a homogeneous electric field to guide the ions
from the reaction zone through the entrance window of the
channeltron. The extraction electric field pulses of 100~V/cm
amplitude and 250 ns duration were applied to the upper plate. The
lower plate was grounded and had a 6~mm diameter opening covered
by a mesh with the transmittance of 70\%. The ions that passed
through the mesh were accelerated by the electric field of the
channeltron to energies of about 3.5~keV.

Single ion output pulses of the channeltron were amplified,
discriminated, and registered by two gated counters. The
measurements were performed in the pulse counting regime, keeping
the frequencies of the detected ion signals much lower (0.2-1~kHz)
than the 5~kHz repetition rate of laser pulses. Less than one ion
per laser shot was detected on average. The measured frequencies
of ion signals were determined as the total number of ions
detected during the measurement time of 10 s, i.e., signals were
counted for 50000 laser pulses. In order to ensure the single-ion
counting regime, the intensity of the laser driving the second
excitation step was attenuated with calibrated neutral density
filters by a factor of 10 to 100.

Our experimental study concentrated on measurements of relative
dependences of ionization rates on the principal quantum number
\textit{n} of Rydberg states, and did not require precise absolute
values of the number density $\mathrm{n}_{3S}$ of the ground-state
Na(\textit{3S}) atoms in the atomic beam. The number density
$\mathrm{n}_{3S}$ in the reaction zone was calculated using a
recent and reliable formula by Browning and Potter
\cite{BrowningPotter}, and it was estimated to be
$\mathrm{n}_{3S}=(5 \pm 1) \times 10^{10}$~$\mathrm{cm}^{-3}$ at
the oven temperature of $T=(635 \pm 2)$~K. Monitoring of the
fluorescence on the saturated resonance transition showed that the
atomic number density was almost constant during the experiments.

The time sequence of excitation and detection pulses is
illustrated in figure~\ref{Fig10}. The first electric field pulse
cleaned the reaction zone from the undesirable atomic and
molecular ions, arisen from photoionization of Rydberg atoms by
laser radiation and photoassociative ionization of sodium
3\textit{P} atoms \cite{Boulmer3p, Burkhardt1985}, respectively.

After the second electric field pulse, the registered Na$^+$ and
Na$_2^+$ ion signals resulted from ionization occurring in the
reaction zone during the time interval $t_2-t_1=1.8$~$\mu$s
between the two extraction electric field pulses. This time is
comparable with the lifetimes of Rydberg states; therefore, time
evolution of ionization processes must be analyzed. The main
processes leading to the production of Na$^+$ ions are
Penning-type ionization (PI) and photoionization by BBR. The
Na$_2^+$ ions can be created only in the associative ionization
(AI). Associative ionization is the simplest two-body collision,
leading to formation of a chemical bond. A Rydberg atom
A(\textit{nL}) collides with the ground-state atom A in the
reactions:
\begin{eqnarray}
&\mathrm{A}\left({nL}\right) + \mathrm{A} \to \mathrm{A}_2^+  +
e^- & \qquad \mathrm{(AI)} \nonumber
\\\nonumber &\mathrm{A}\left( {nL} \right) + \mathrm{A} \to \mathrm{A}^+  + \mathrm{A} + e^-
&\qquad \mathrm{(PI)}\nonumber
\end{eqnarray}
\noindent A contribution of the collisions with background gases
can be safely disregarded. We have verified experimentally that
the variation of background pressure within the range of 5$ \times
10^{-7}\le \mathrm{P} \le 1 \times 10^{-6}$~Torr did not affect
the measured Na$^+$ and Na$_2^+$ signals by more than 5\%. Under
such conditions, the rate equations describing the evolution of
the number of Na$^+$ and Na$_2^+$ ions following the laser
excitation at time \textit{t}=0 are [see equation~(\ref{eq31})]

\begin{equation}
\label{eq36} \left\{{{\begin{array}{*{20}l} {\frac{\mathstrut
{\mbox{\textit{d}Na}^{ +} \left( {t}
\right)}}{{\mbox{\textit{dt}}}} = k_{PI} N_{nL} \left( {t}
\right)\,\mathrm{n}_{3S} + W_{BBR} N_{nL} \left( {t} \right);}\\
\\

{\frac{{\mbox{\textit{d}Na}_{{2}}^{ +} \left( {t}
\right)}}{{\mbox{\textit{dt}}}} = k_{AI} N_{nL} \left( {t}
\right)\,\,\mathrm{n}_{3S}.}
\\
\end{array}} } \right.
\end{equation}

\noindent Here $N_{nL} \left( {t} \right) \approx N_{nL} \left(
{0} \right)\,\,\mathrm{exp}\left[ { - t/\tau _{eff}}  \right]$ is
the time-dependent number of Na(\textit{nL}) Rydberg atoms in the
reaction zone, $\mathrm{n}_{3S} = 5 \times 10^{10} \quad
\mathrm{cm}^{-3}$ is the number density of ground state atoms,
$k_{AI}$ and $k_{PI}$ are rate constants of associative and
Penning ionization in Na($nL$)+Na($3S$) collisions. The initial
number of Rydberg atoms, $N_{nL}$(0), created during laser
excitation can be written as

\begin{equation}
\label{eq37}
N_{nL} \left( {0} \right) = N_{3P} W\left( {3P_{3/2} \to nL} \right),
\end{equation}

\noindent where $N_{3P}$ is the average number of atoms in the
$3P_{3/2}$ state during the yellow-laser pulse, and $W\left(
{3P_{3/2} \to nL} \right)$ is the probability of excitation of the
Na($3P_{3/2}$) atoms to the \textit{nL} state by a single
blue-laser shot.

The effective lifetime $\tau _{eff}$ describing the decay of
Rydberg states in equation~(\ref{eq10}) is determined by the
spontaneous lifetime and the rate of other processes depleting the
laser excited Rydberg state. These include BBR induced transitions
between Rydberg states, BBR induced photoionization, and
collisional quenching.

The depletion of Rydberg states with \textit{n}=8-20 by
collisional ionization is negligible at the atom density used in
our experiment. According to our estimates, the rate of
associative ionization, $k_{AI} \mathrm{n}_{3S}$, does not exceed
50 s$^{-1}$ and is therefore much smaller than the spontaneous
decay rates, which range from $10^5$ to $10^6$ s$^{-1}$ for the
studied Rydberg states. The rate of PI, $k_{PI} \mathrm{n}_{3S}$,
is expected to be below 10~s$^{-1}$ for $n \sim $20, and close to
zero for lower \textit{n}. Comparing the PI rate with the direct
BBR photoionization rate $W_{BBR}$, one can see that Na$^+$ ions
are produced mainly via BBR photoionization. As will be shown
below, this background ionization process can be favorably
exploited for the determination of absolute AI rate constants.

With the above considerations in mind, the solution of
equations~(\ref{eq36}) can be written as

\begin{equation}
\label{eq38}
\left\{ {{\begin{array}{*{20}c}
 {\mathrm{Na}^{ +}  = N_{nL} \left( {0} \right)\;W_{BBR} \;t_{eff}}  \\
 {\mathrm{Na}_{2}^{ +}  = N_{nL} \left( {0} \right)k_{AI} \mathrm{n}_{3S} \;t_{eff}}  \\
\end{array}} } \right.
\end{equation}

\noindent where $t_{eff}$ is the effective time of interaction
that takes into account the short radiative lifetimes of Rydberg
states, determined by equation~(\ref{eq33})

Equations~(\ref{eq38}) can be used for a direct measurement of
$k_{AI}$ and $W_{BBR}$ values, provided $N_{nL}(0)$ is known. The
only reliable method to measure $N_{nL}(0)$ is the SFI technique.
Unfortunately, SFI method is difficult to apply to Rydberg states
with low \textit{n}, since it requires too strong electric field
($\sim $30~kV/cm for $n \sim 10$).

On the other hand, we were interested mainly in relative
measurements of $W_{BBR}$ for various \textit{n}. Therefore we
could use a normalization procedure for $N_{nL} (0)$ based on
numerically calculated excitation probabilities $W\left( {3P_{3/2}
\to nL} \right)$. Since the $3S_{1/2} \to 3P_{3/2}$ transition was
saturated, $N_{nL} (0)$ depended only on the respective transition
moments and power of the blue laser. In the absence of saturation
at the second excitation step (this was the case for our
experiments), the probability of excitation of Rydberg states from
the $3P_{3/2}$ state can be written as

\begin{equation}
\label{eq39} W\left( {3P_{3/2} \to nL} \right) = C_{L} \cdot I_{b}
\cdot R^{2}\left( {3P_{3/2} \to nL} \right),
\end{equation}

\noindent where $I_{b}$ is the power of the blue laser, $R\left(
{3P_{3/2} \to nL} \right)$ is the radial part of the transition
dipole moment, and $C_{L}$ is a normalization constant which
depends on \textit{L} and is proportional to the square of angular
part of the matrix element. $W\left( {3P_{3/2} \to nL} \right)$
falls as $n_{eff}^{-3}$ for high Rydberg states, but for the
states with $n\sim 10$ this scaling low does not work well. We
have revealed this fact in our numeric calculations of $R\left(
{3P_{3/2} \to nL} \right)$ for the $3P_{3/2} \to nS, nD$
transitions, and therefore used the numerical data in subsequent
measurements instead of the scaling law.

In order to compare the absolute signals due to BBR and
collisional ionization of \textit{nS} and \textit{nD} states, it
is necessary to know also the ratio $C_D/C_S$. The analysis of
angular parts of the transition matrix elements, taking into
account the hyperfine structure, has shown that for excitation
with linearly polarized light in the first and the second
excitation steps, the ratio $C_D / C_S$ may vary from
approximately 1.6 (if there is no collisional, radiative, or
magnetic field mixing of the magnetic sublevels) to 2 (if the
sublevel mixing is complete). For excitation by non-polarized
light, the ratio always equals to 2 regardless the degree of level
mixing. Finally, we find that the ratio $W\left( {3P_{3/2} \to nD}
\right)/W\left( {3P_{3/2} \to nS} \right)$ may vary between the
3.5 and 5.

%==============================================================================================
%==============================================================================================
%=======================*************FIGURE 15**********=========================================
%===============================================================================================
\begin{figure}

\begin{center}
\includegraphics[width=13cm]{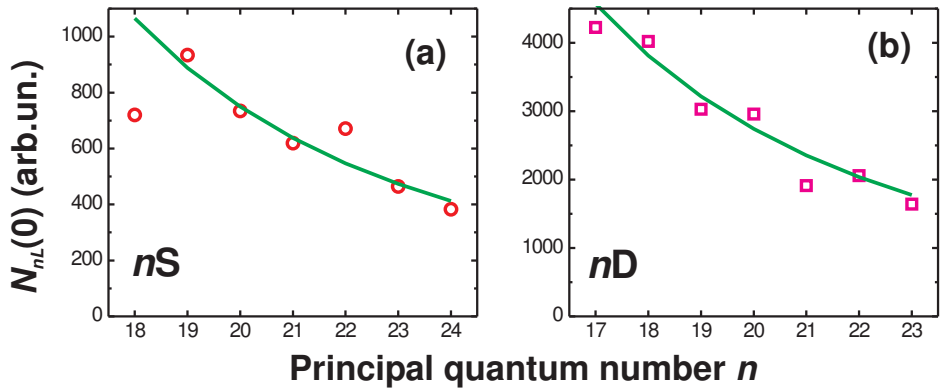}% Here is how to import EPS art
\caption{\label{Fig15} Relative probabilities of laser excitation
of sodium Rydberg states: (a) \textit{nS} states; (b) \textit{nD}
states. Open circles and squares - experiment, solid curves -
theory.}
\end{center}
\end{figure}

%==============================================================================================

In principle, one could normalize the ion signals measured for
different \textit{nL} states using the calculated probabilities
$W\left( {3P_{3/2} \to nL} \right)$ and measuring only the power
$I_{b}$ of the blue laser radiation and equation~(\ref{eq39}).
However, the applicability of such normalization may be
complicated by technical imperfections of the blue laser. Since
the linewidth of this laser (10~GHz) was much larger than the
widths of the absorption profiles at the second excitation step
($\sim $500~MHz Doppler broadening), variations of the spectral
density of laser radiation could affect the probability of
excitation even if $I_{b}$ would be kept constant. Therefore we
had to verify experimentally the applicability of normalization by
equation~(\ref{eq39}). As discussed above, the only reliable way
to measure the number of Rydberg atoms was to apply the SFI
technique. For this purpose, we built a high-voltage generator
yielding pulses with rise time of 1~$\mu$s and amplitude of up to
8 kV. This allowed us to field-ionize Rydberg states with $n \ge
17$. The SFI signals were detected at a 1~$\mu$s delay with
respect to the laser pulse, i.e., the measured SFI signal was:

\begin{equation}
\label{eq40} S_{SFI} \sim N_{nL} \left( {0}
\right)\,\mathrm{exp}\left( { - 1\;\mu\mathrm{s}/\tau _{eff}}
\right).
\end{equation}

\noindent Equation~(\ref{eq40}) was used to derive $N_{nL} (0)$
from the measured SFI signals and the calculated values of $\tau
_{eff}$, which were published in Ref.~\cite{Lifetimes}.
Figure~\ref{Fig15} shows the measured $N_{nL}(0)$ dependences on
the principal quantum number \textit{n} for \textit{nS} and
\textit{nD} states. These data are normalized over $I_{b}$,
because it varied as the blue laser frequency was tuned to
resonance with different \textit{nL} states. The solid curves are
the approximations made using equation~(\ref{eq39}). It is seen
that experimental points have noticeable deviations from theory
although the general trend is correct. These deviations may be
explained by the variations of spectral density of the blue laser
light. We concluded that equation~(\ref{eq39}) can be used for the
normalization of $N_{nL} (0)$, but at a price of limited accuracy.
We also find from figure~\ref{Fig15} that average ratio
$W(3P_{3/2} \to nD)/W(3P_{3/2} \to nS)$ was close to 3.5. Hence,
no considerable mixing of the magnetic sublevels took place during
laser excitation, and the ratio $C_D/C_S$ was close to 1.6.

%==============================================================================================
%==============================================================================================
%=======================*************FIGURE 16**********=========================================
%===============================================================================================
\begin{figure}

\begin{center}
\includegraphics[width=13cm]{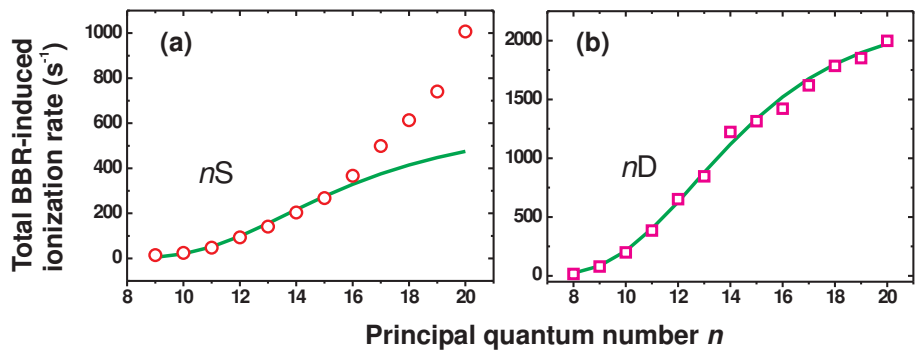}% Here is how to import EPS art
\caption{\label{Fig16} Total BBR induced ionization rates for (a)
\textit{nS} states and (b) \textit{nD} sodium Rydberg states. Open
circles and squares - experiment, solid lines - theory.}
\end{center}
\end{figure}

%==============================================================================================

Experimental and theoretical data on the total Na BBR ionization
rates at $T$=300~K are compared in figure~\ref{Fig16}. The solid
curves are the theoretical values of $W_{BBR}^{tot}$ (see
Section~2). The squares and circles are the corresponding
experimental values $W_{BBR}^{exp}$ obtained with
equations~(\ref{eq38}) and (\ref{eq39}) [in equation~(\ref{eq38})
the value of $W_{BBR}$ was replaced with $W_{BBR}^{tot} $].
Experimental data were averaged over 5 measurements. The
normalization coefficient \textit{C}$_{D}$\textit{} in
equation~(\ref{eq39}) was the only free parameter whose absolute
value was adjusted to fit experiment and theory. A remarkably good
agreement between the experimental and theoretical data was found
for \textit{n}D states [figure~\ref{Fig16}(a)]. At the same time,
the data for \textit{nS} states, obtained with measured earlier
ratio $C_D/C_S=1.6$, exhibit considerable discrepancies for states
with higher \textit{n} [figure~\ref{Fig16}(b)], while the
agreement for states with lower \textit{n} is much better. The
values of $W_{BBR}^{exp} $ exceed the values of $W_{BBR}^{tot}$ by
2.1 times for \textit{n}=20, and the shape of the experimental
\textit{n} dependence was significantly different from the
theoretical one.

One possible explanation of  anomaly for \textit{nS} states is
related to their specific orbit that penetrates into the atomic
core. The penetration causes a strong interaction between the
Rydberg electron and the core, e.g., due to core polarization
\cite{Aymar}. This results in a large quantum defect and a Cooper
minimum in the photoionization cross-sections. This assumption is
supported by the good agreement of theory and experiment for the
hydrogen-like \textit{nD} states, which have a small quantum
defect and almost non-penetrating orbits.

%===============================================================================================
%===============================================================================================
%===============================================================================================

\subsection{Application to measurements of collisional
ionization rates}

%===============================================================================================
%===============================================================================================

Since SFI technique was not applicable for direct determination of
$N_{nL}$(0) for \textit{n}=8-17, and the use of
equation~(\ref{eq40}) was seen to be somewhat inadequate in our
experiment, we had to find a way to eliminate this value in the
measurements. We decided to measure the ratio \textit{R} of atomic
and molecular signals derived from equations~(\ref{eq38}):

\begin{equation}
\label{eq41} R = \frac{{\mathrm{Na}_{2}^{ +} } }{{\mathrm{Na}^{ +}
}} = \frac{{k_{AI} \mathrm{n}_{3S} }}{{W_{BBR}^{tot}} }.
\end{equation}

\noindent This ratio is independent of the values of $N_{nL}$(0),
$\tau _{eff}$ and $t_{eff}$. Thus, the rate constant of the AI
process can be directly obtained from the measured ratio of the
Na$_2^+$ and Na$^+$ signals:

\begin{equation}
\label{eq42} k_{AI} = \frac{{\mathrm{Na}_{2}^{ +} }
}{{\mathrm{Na}^{ +} }} \cdot \frac{{W_{BBR}^{tot}
}}{{\mathrm{n}_{3S}} }.
\end{equation}

\noindent The BBR ionization rates $W_{BBR}$ became to be key
values necessary for the determination of the AI rate constants.
Therefore, an accuracy with which the $W_{BBR}$ values are known
determines the accuracy of the experimental $k_{AI}$ values
obtained with equaiton~(\ref{eq42}).

In our experiments associative ionization rate constants were
measured separately in single and crossed Na atomic beams at
temperatures of \textit{T}=635~K (single beam) and
\textit{T}=600~K (crossed beams). The results of these
measurements were published in our previous works \cite{PaperI,
PaperIII}.

%===============================================================================================
%===============================================================================================
%===============================================================================================
\subsection{Experimental studies of ultracold plasma}
%===============================================================================================
%===============================================================================================
%===============================================================================================
The mechanism of formation of an ultracold plasma from a dense
sample of cold Rydberg atoms was briefly described in the
Introduction.  Roughly 2/3 of Rydberg atoms are converted into a
plasma, while the remaining atoms decay to low-lying states, thus
keeping the energy balance.

Spontaneous evolution of cold Rydberg atoms into ultracold plasma
was first observed by Robinson et al. \cite{Robinson2000}.
Experiments were performed with Rb and Cs Rydberg atoms held in a
magneto-optical trap (MOT). The cloud of cold atoms had a
temperature of 300~$\mu$K in the case of rubidium, and 140~$\mu$K
in the case of cesium. The atoms were excited from the
$5P_{3/2}$(Rb) or $6P_{3/2}$(Cs) states to the Rydberg states by
radiations of the pulsed dye lasers. Untrapped room-temperature
atoms were also excited into Rydberg states, which made 1\%
contribution to the total number of excited Rydberg atoms. At
delay time $t_{d}$ after the laser pulse, a rising voltage pulse
was applied to the parallel electric-field plates surrounding the
excitation volume. The time $t_{d} $ was varied in the interval
0-50~$\mu$s. The rising pulse first frees electrons bound to the
plasma, then ionizes Rydberg atoms and drives electrons (or ions,
depending on the polarity) to a microchannel-plate detector~(MCP).
The time resolved ionization signals were studied. A plasma
signal, which came before the field ionization pulses, was
observed even at delay times $t_{d} = 20~\mu$s, which observation
demonstrated that Rydberg atoms had evolved into a plasma.

Later, Gallagher et al.~\cite{Ultracold5, Ultracold6} studied the
role of dipole-dipole interaction for the ionization of ultracold
Rydberg gas. It has been shown that for Rydberg states with $n<40$
BBR and collisions are the predominant sources of initial
ionization, but for higher states ionization is caused mostly by
the resonant dipole interaction of Rydberg atoms. These results
show that accurate calculations and experimental measurements of
the rates of BBR-induced and collisional ionization are of great
importance for contemporary studies of the formation of ultracold
plasma.

\section{Conclusion}

We have calculated the total BBR-induced ionization rates of
\textit{nS}, \textit{nP} and \textit{nD} Rydberg states of all
alkali-metal atoms for principal quantum numbers \textit{n}=8-65
at the ambient temperatures of 77, 300 and 600~K. Our calculations
take into account the effect of BBR-induced mixing of Rydberg
states and their field ionization by extracting electric field
pulses. Useful analytical formulas have been derived, which allow
for quick estimation of ionization rates and their dependences on
the principal quantum number \textit{n}. The numerical results are
in a good agreement with our recent experiment data on Na
\textit{nS} and \textit{nD} states, except for \textit{nS} states
with $n>15$, which is most probably associated with the Cooper
minimum in the photoionization cross-section.

The obtained results show that BBR-induced redistribution of
population over Rydberg states and their field ionization by
extracting electric fields affect both the magnitudes of the total
ionization rates and shapes of their dependences on the principal
quantum number. This suggests that these processes are important
and cannot be ignored in the calculations and measurements of BBR
ionization rates. Equations~(\ref{eq31})-(\ref{eq35}), as well as
the analytical formulas~(\ref{eq27}) and (\ref{eq30}), can be used
to calculate the total ionization rates $W_{BBR}^{tot} $ under
particular experimental conditions. The numerical results
presented in figures~\ref{Fig3}-\ref{Fig13} may be helpful to the
analysis of ionization signals measured in experiments on
collisional ionization and spontaneous formation of ultracold
plasma, since BBR-induced ionization is the main source of atomic
ions. New experimental data for alkali-metal Rydberg atoms in a
broader range of principal quantum numbers would be of interest
for the further improvement of theory, especially for the
non-hydrogen-like states.

The set of results obtained by us constitutes the first systematic
study of BBR-induced ionization of all alkali-metal Rydberg atoms.
It may be helpful in the analysis of the mechanism of ultracold
plasma formation in different experimental conditions
\cite{Robinson2000, LiNoele2004}.

\section{Acknowledgments}

This work was supported by the Russian Academy of Science, Dynasty
Foundation,  EU FP6 TOK Project LAMOL (Contract
MTKD-CT-2004-014228), Latvian Science Council and European Social
Fund.

\end{document}